\documentclass[10pt,a5paper,fleqn]{article}

\input{preamble.tex}

\usepackage[mathlines]{lineno}

\author{Marcelo Righi\footnote{Corresponding author. We are grateful for the financial support of CNPq (Brazilian Research Council) projects numbers 302614/2021-4 and 401720/2023-3 and FAPERGS project number 25/2551-0000969-3.}\\ \small{ \href{mailto:marcelo.righi@ufrgs.br}{marcelo.righi@ufrgs.br}} \and Eduardo Horta\\\small{\href{mailto:eduardo.horta@ufrgs.br}{eduardo.horta@ufrgs.br}}
\and Marlon Moresco\\\small{\href{mailto:marlon.moresco@ufrgs.br}{marlon.moresco@ufrgs.br}}}

\date{\scshape{Universidade Federal do Rio Grande do Sul, Brazil}}

\title{\scshape Set Risk Measures}

\begin{document}
\frenchspacing
\maketitle

\begin{abstract}

We introduce set risk measures (SRMs), real-valued maps defined on the family of non-empty closed bounded sets of essentially bounded random variables. SRMs extend traditional scalar risk measures by assigning a single capital requirement to an entire set of positions. We develop an axiomatic framework for SRMs, adapting classical properties such as monotonicity, translation invariance, convexity, and positive homogeneity to set arithmetic. The main technical contribution is a dual representation of convex SRMs through the \strict{} topology and regular $\tau$-additive unit-mass measures. We also characterize worst-case SRMs and present examples related to systemic risk, Knightian uncertainty, and preference representations.

\vspace{.1in}
\end{abstract}

\clearpage
\begingroup
\setlength{\LTleft}{0pt}
\setlength{\LTright}{0pt}
\setlength{\tabcolsep}{0pt}
\newlength{\notationlabelwidth}
\newlength{\notationgap}
\setlength{\notationlabelwidth}{.33\linewidth}
\setlength{\notationgap}{1.5em}
\renewcommand{\arraystretch}{1.15}
\newcommand{\notationrow}[2]{#1 & #2\\}
\begin{longtable}{@{}>{\raggedright\arraybackslash}p{\notationlabelwidth}@{\hspace{\notationgap}}>{\raggedright\arraybackslash}p{\dimexpr\linewidth-\notationlabelwidth-\notationgap\relax}@{}}
\multicolumn{2}{@{}p{\linewidth}@{}}{\large\bfseries Index of Notation}\\[.75\baselineskip]
\endfirsthead
\multicolumn{2}{@{}p{\linewidth}@{}}{\large\bfseries Index of Notation}\\[.75\baselineskip]
\endhead
\notationrow{$\ba$}{The space of all finitely additive signed measures on $(\Omega,\sigmafield{F})$ that are absolutely continuous with respect to $p$ and with finite total variation}
\notationrow{$\ba_{+}$}{The set of elements $q\in\ba$ such that $q(A)\ge0$ for all $A\in\sigmafield{F}$}
\notationrow{$\ba_{1,+}$}{The set of elements $q\in\ba_{+}$ such that $q(\Omega) = 1$}
\notationrow{$U$}{The closed unit ball in $(\ba,\Vert\cdot\Vert_{\textsc{tv}})$}
\notationrow{$U_{+}$}{The non-negative elements in $U$, i.e., $U_{+} = U\intersection \ba_{+}$}
\notationrow{$\sigmafield{A}_U$}{The \emph{algebra} generated by the open subsets of $U$}
\notationrow{$\bare$}{The set of regular, signed finitely additive measures on $(U,\sigmafield{A}_U)$ with finite variation}
\notationrow{$\bare_+$}{The set of elements $\mu\in\bare$ such that $\mu(A)\ge0$ for all $A\in\sigmafield{A}_U$}
\notationrow{$\bare_{1,+}$}{The set of elements $\mu\in\bare_+$ such that $\mu(U) = 1$}
\notationrow{$\Pbare$}{The set of elements $\mu\in\bare_{1,+}$ such that $\mu(\ba_{1,+}) = 1$}
\notationrow{$\sigmafield{B}_{U}$}{The $\sigma$-field generated by the open subsets of $U$}
\notationrow{$\tauare$}{The set of regular, signed $\tau$-additive measures on $(U,\sigmafield{B}_U)$ with finite total variation}
\notationrow{$\tauare_+$}{The set of elements $\mu\in\tauare$ such that $\mu(A)\ge0$ for all $A\in\sigmafield{B}_{U}$}
\notationrow{$\tauare_{1,+}$}{The set of elements $\mu\in\tauare_{+}$ such that $\mu(U) = 1$}
\notationrow{$\Ptau$}{The set of elements $\mu\in\tauare$ such that $\mu(A)\ge0$ for all $A\in\sigmafield{B}_{U}$, $\mu(U) = 1$, and $\mu(\ba_{1,+}) = 1$}
\notationrow{$\alpha,\gamma,\varepsilon,\lambda$}{Real scalars}
\notationrow{$x,y,z,\dots$}{Random variables}
\notationrow{$X,Y,Z,\dots$}{Sets of random variables}
\notationrow{$\rvsetset{X},\rvsetset{Y},\rvsetset{Z},\dots$}{Collections of sets of random variables}
\notationrow{$p,q,\dots$}{Finitely additive signed measures on $(\Omega,\sigmafield{F})$}
\notationrow{$P,Q,\dots$}{Sets of measures}
\notationrow{$\E{q}{x}$}{Integral of the random variable $x$ with respect to the measure $q$}
\notationrow{$\E{Q}{X}$}{$\sup_{(q,x)\in Q\times X}\E{q}{x}$}
\notationrow{$\CB[M]$}{Space of all non-empty, closed, and bounded subsets of a metric space $M$}
\notationrow{$\CBC(M)$}{Space of all non-empty, closed, bounded, and convex subsets of a metric vector space $M$}
\notationrow{$C_b(M)$}{The vector lattice of all functions $f\colon M\to\R$ that are continuous and bounded, defined on a topological space $M$}
\notationrow{$\Rbar$}{The extended real line, i.e., $\Rbar = \R\cup\lbrace-\infty,\infty\rbrace$}
\end{longtable}
\endgroup

\clearpage

\section{Introduction}

Risk measures have become a cornerstone of Mathematical Finance, particularly following the influential work of \citet{Artzner1999} on coherent risk measures---see \citet{Follmer2016} for a comprehensive review. Traditionally, the theory, as summarized in \citet{Delbaen2002}, considers risk measures as real-valued functionals defined on the domain $L^\infty$ of bounded random variables. However, the increasing complexity of modern financial markets demands more versatile tools capable of addressing multivariate and systemic aspects of risk, which motivates extending the traditional framework.

In this paper, we introduce the concept of a \textit{set risk measure} (SRM), defined as a real-valued map whose domain, $\CB$, comprises all non-empty, closed, and bounded subsets of $L^\infty$. Our rationale mirrors the foundational reasoning behind monetary risk measures built from acceptance sets, aiming to quantify the minimum capital required to render an entire set of financial positions acceptable. The central technical challenge of this extension is that $\CB$ is not a linear space, limiting the direct application of conventional analytical tools. Our approach leverages advanced topological, algebraic, and order-theoretic methods to overcome these limitations.

Spaces like $\CB$ belong to a broader class known as \emph{hyperspaces}, equipped with hypertopologies. We work primarily with the Hausdorff distance topology, under which the canonical mapping $x\mapsto \lbrace x\rbrace$ is a homeomorphism onto its image \citep{Lucchetti1994}. This structural property reveals a fundamental distinction from the traditional domain $L^\infty$: there is no straightforward reduction from SRMs to standard scalar risk measures. This observation emphasizes that our set-based approach genuinely extends beyond the traditional scalar-to-scalar setup.

Our primary contribution is thus establishing this novel framework of ``set-to-scalar'' risk measures, generalizing existing univariate and multivariate theories. While numerous domains beyond $L^\infty$ have been considered, including $L^{\lambda}$ spaces \citep{Kaina2009}, Orlicz spaces \citep{Cheridito2009,Gao2018}, and spaces of probability distributions \citep{Frittelli2014}, these remain fundamentally scalar and do not capture multivariate intricacies. Even multivariate extensions found in \citet{Burgert2006,Burgert2008,Rueschendorf2006,Ekeland2011,Ekeland2012}, and others, typically handle finite-dimensional vectors rather than general sets of random variables. SRMs are designed precisely to accommodate this greater generality, with systemic risk as a natural motivating application.

Our framework also differs from existing ``vector-to-set'' approaches such as set-valued risk measures studied by \citet{Jouini2004,Hamel2010,Hamel2011,Ararat2017}, and from risk measures for random sets as developed by \citet{Molchanov2016} and \citet{Molchanov2021}. While these methodologies either produce set-valued outcomes or assess sets in a fundamentally different manner, our approach specifically quantifies the overall acceptability of a set as a single scalar, capturing the essential financial intuition of capital adequacy in a straightforward numerical form.

Additionally, our theory is distinct from risk measures defined over spaces of stochastic processes \citep{Cheridito2004,Cheridito2005,Frittelli2006}, which are tailored to time-dependent structures. The generality of our SRM framework allows a broader class of set-based financial positions to be assessed directly, without imposing restrictive structures inherent to stochastic processes or Banach lattice topologies.

The closest work to ours is \citet{Fadina2023}, where the input is a random variable together with a set of probability measures. Their framework also imposes set-monotonicity and WC-boundedness, in a sense closely related to the properties considered here. Nonetheless, the present paper adopts a more general viewpoint. While these authors analyze the risk of a single position when there is uncertainty about the underlying probability measure, i.e., risk is assessed with respect to a set of models, our framework defines risk directly on entire sets of positions, independently of whether model uncertainty is present. In this sense, SRMs provide a functional-analytic generalization: the case in which positions are fixed and only probabilities vary appears as a particular instance within the broader space of random-variable sets.

The paper is organized as follows. In \Cref{sec:SRM}, we propose an \emph{axiomatic scheme} for set risk measures, extending classical scalar-risk axioms and introducing additional properties that arise naturally from set operations. We also establish the connection between SRMs and their induced acceptance sets, in direct analogy with the classical construction.

In \Cref{sec:dual}, we present the main technical result: an axiomatic dual representation of convex SRMs in terms of regular, $\tau$-additive (for nets) measures on the unit ball of $\ba$, the dual of $L^\infty$. The key observation is that the space of non-empty, closed, bounded, and convex subsets of $L^\infty$ embeds naturally as a cone in the space of bounded continuous functions on this dual unit ball. This embedding permits a representation in terms of \emph{bona fide} probability measures. Bounded Wijsman convergence is then used to connect this set-valued geometry with the \strict{} topology on the embedded space This representation is particularly valuable from both theoretical and practical perspectives, enhancing interpretability and computational tractability.

In \Cref{sec:WC}, we analyze \emph{worst-case risk measures}, namely SRMs that evaluate a set by the supremum of the risks of its individual elements. This class has attracted considerable attention because of its transparent financial interpretation and its relevance for robust risk management \citep{Follmer2002,Laeven2013,Ang2018,Bartl2018,Bellini2018}. We characterize when an SRM is of worst-case type and refine the associated continuity properties and dual representations.

Finally, \Cref{sec:examp} offers illustrative examples highlighting the practical utility of our theoretical framework, demonstrating its relevance for systemic risk evaluation, portfolio optimization, and decision-making under uncertainty. Through these concrete cases, we underscore the flexibility and potential of set risk measures in addressing contemporary challenges in financial risk management.

\section{Set risk measures}\label{sec:SRM}

We consider a probability space $(\Omega,\sigmafield{F},p)$, and denote by $L^0\coloneqq L^0(\Omega,\sigmafield{F},p)$ and $L^{\infty}\coloneqq L^{\infty}(\Omega,\sigmafield{F},p)$ the spaces of (equivalence classes under $p\text{-a.s.}$ equality of) finite and essentially bounded random variables, respectively. Throughout, lowercase $p,q$ denote measures, while uppercase $P,Q$ denote sets of such measures. Similarly, lowercase $x,y,z$ denote random variables, while $X,Y,Z$ denote sets of such random variables. All equalities and inequalities are understood in the $p$-almost sure sense; in particular, constant random variables \emph{are} real numbers. Unless stated otherwise, $L^\infty$ is equipped with its norm topology. We write $1_A$ for the indicator of $A\in\sigmafield{F}$, denote by $\cl(X)$ the \defin{closure} of $X$, and by $\conv(X)$ the \defin{closed convex hull} of $X\subseteq L^\infty$. In both cases, closure is taken with respect to the supremum norm.

Consider the power set $\Pow(L^\infty)$, i.e., $X\in \Pow(L^\infty)$ if and only if $X\subseteq L^\infty$. We begin by endowing $\Pow(L^\infty)$ with a metric structure, as follows: For any pair $X,Y\subseteq L^\infty$ of non-empty sets, define the \defin{infimal set distance} between the two sets as $\di(X,Y) = \inf_{x\in X,y\in Y}\lVert x-y\rVert_{\infty}$ (note that $\di$ is not a metric). We set $\di(X,Y) = \infty$ if either $X$ or $Y$ is empty. The \defin{Hausdorff distance} on $\Pow(L^\infty)$ is then defined through
\begin{equation*}
d_H(X,Y) = \max\left\lbrace \sup_{x\in X}\di(\singleton{x}, Y), \sup_{y\in Y}\di(\singleton{y}, X)\right\rbrace,\qquad X,Y\in\Pow(L^\infty).
\end{equation*}
In what follows, the notation $\Hlim X_\alpha = X$ means $d_H(X_\alpha, X)\to0$, where $(X_\alpha : \alpha \in A)$ is a net in $\Pow(L^\infty)$. We now recall some properties of $d_H$ that will be used without further mention \citep[see][for details]{Beer1993}.

\begin{enumerate}[label = \arabic*., noitemsep]

\item $d_H(X,Y) = \sup_{x\in L^\infty}\left| \di(\singleton{x},X)-\di(\singleton{x},Y)\right|$.

\item If $Y = \singleton{y} $ is a singleton, then $d_H(X,Y) = \sup_{x\in X}\lVert x-y\rVert_\infty$.

\item $d_H(X,Y)\in[0,\infty]$. If both sets are bounded, then $d_H$ is guaranteed to be finite. Moreover, $d_H(X,Y) = 0$ if and only if $\cl(X) = \cl(Y)$.

\item $\di(\singleton{x},X) \leq \di(\singleton{x},Y)+d_H(X,Y)$ for any $x\in L^\infty$.

\item If $\interior(X\intersection Y)\neq\varnothing$, then there is a $\lambda > 0$ such that $Y\intersection Z\neq\varnothing$ for every $Z\subseteq L^\infty$ satisfying $d_H(X,Z)<\lambda$.

\item $d_H$ is an extended pseudometric on $\Pow(L^\infty)$.
\end{enumerate}

We write $\CB$ for the hyperspace of all non-empty, closed, and bounded subsets of $L^\infty$, and $\CBC(L^\infty)$ for the subclass of convex elements of $\CB$. Importantly, the Hausdorff distance $d_H$ induces a complete metric on $\CB$ \citep[see][Theorem 3.2.4, p.~87]{Beer1993}.

We thus focus initially on the complete metric space $(\CB,d_H)$. The map $(X,Y)\mapsto X\cup Y$ is continuous on $\CB$, and $(X,Y)\mapsto\conv\left(X\cup Y\right)$ is continuous on $\CBC(L^\infty)$. In addition, note that the collection of finite sets in $L^\infty$ is dense in the collection of non-empty compact sets in $L^\infty$.

We next introduce the algebraic operations on $\Pow(L^\infty)$. Although these operations depend on the topology of $L^\infty$ through closure, they are used here as algebraic operations on sets. The \defin{Minkowski sum} is defined by $X+Y = \cl\left(\lbrace x+y\colon x\in X, y\in Y\rbrace \right)$, and we write $X+\singleton{x}$ and $X+x$ interchangeably when no confusion can arise. This operation is commutative and associative, with $\singleton{0}$ as the identity element. Scalar multiplication is defined by $\lambda X = \lbrace \lambda x \colon x\in X\rbrace $ for $\lambda \in\R$, and $X-Y = X+(-Y)$. We also set $X + \varnothing = \varnothing$ and $\lambda \varnothing = \varnothing$. Despite this notation, these operations do not make $\Pow(L^\infty)$ into a linear space, since Minkowski addition lacks cancellation.

Since $d_H(X+\singleton{\lambda},Y+\singleton{\lambda}) = d_H(X,Y)$ and $d_H(\lambda X, \lambda Y) = |\lambda |d_H(X,Y)$ for every $\lambda \in\R$, translations and scalar multiplication are continuous in the corresponding Hausdorff topologies. Moreover, $\lambda (X+Y) = \lambda X+\lambda Y$ and $\lambda_1(\lambda_2X) = (\lambda_1\lambda_2)X$, although in general $\lambda_1X+\lambda_2X \ne (\lambda_1+\lambda_2)X$. Thus $\Pow(L^\infty)$ is a \defin{conlinear} space, and $\CB$ is a conlinear subspace of $\Pow(L^\infty)$ \citep[for a review on conlinear spaces, see][]{schrage2009set,hamel2005variational}. With a mild abuse of terminology, we use vector-space notions such as convex set, convex hull, and linear function whenever the relevant operations are well-defined. Finally, define the \defin{extended gauge} on $\Pow(L^\infty)$ by $X \mapsto\lVert X\rVert = \sup_{x\in X}\lVert x\rVert_\infty = \inf \{ \alpha \in \R : p(|x|>\alpha) = 0 \text{~for all~} x\in X \}$. This gauge is subadditive, positively homogeneous, and finite on bounded sets.

A standard preorder in $\Pow(L^\infty)$ is defined as follows: we say that $Y$ \defin{is greater than or equal to} $X$, denoted by $X \leq Y$, if for any $y\in Y$ there is an $x\in X$ such that $x \leq y$, and we write $Y\geq X$ to mean $X\leq Y$. This relation is reflexive and transitive, but not antisymmetric. It is compatible with the algebraic operations on $\CB$: if $X\leq Y$, then $X+Z\leq Y+Z$ for every $Z\in\CB$, and $\lambda X\leq\lambda Y$ for every $\lambda \geq0$. Equivalently, $X\leq Y$ if and only if $X+ L^\infty_{+} \supseteq Y$. Restricted to monotone subsets of ${L^\infty}$, namely sets satisfying $X+L^\infty_{+} = X$, this preorder coincides with the order-complete lattice order induced by set inclusion. Finally, $Y \subseteq X$ implies $Y \geq X$. For further details on this order and its link to preferences, see \citet{kreps1979representation,dekel2001representing}.

\begin{Rmk}\label{rmk: orders}
An alternative order on $\Pow(L^\infty)$ is $X \lesssim Y$, meaning that for every $x\in X$ there exists $y\in Y$ such that $x \leq y$. This order is natural when $x$ represents a loss rather than a profit-and-loss position; see \citet{moresco2023uncertainty}. Indeed, $X \lesssim Y$ if and only if $-Y \leq -X$. Another possibility is to set $X\succeq Y$ if and only if $x\geq y$ for every $(x,y)\in X\times Y$. This is equivalent to $(X-Y)\subseteq L^\infty_{+}$, a convex cone, and also to $\essinf X \geq \esssup Y$; see the definition below. This relation is not a preorder, since reflexivity may fail. It is, however, stronger than ``$\ge$'': $X\succeq Y$ implies $X\ge Y$.
\end{Rmk}

Define the \defin{absolute value} of $X$ by $|X| = \lbrace |x|\colon x\in X\rbrace $. For $X\in \Pow(L^\infty)$, its \defin{essential supremum}, denoted by $\esssup X$, is the almost surely unique random variable $y\colon\Omega\to\R\cup\{+\infty\}$ such that $y\geq x$ for every $x\in X$, and $y\leq z$ for every $z\colon\Omega\to\R\cup\{+\infty\}$ satisfying $z\geq x$ for every $x\in X$. The \defin{essential infimum} is defined by $\essinf X = -\esssup(-X)$. By allowing extended-real-valued random variables, $\esssup X$ is well defined \citep[see Theorem A.37 in][]{Follmer2016}; when $X$ is countable, it may be taken pointwise as $(\esssup X)(\omega) = \sup_{x\in X} x(\omega)$. The essential supremum is finite if and only if all elements of $X$ are bounded above by some $z\in L^0$; it is integrable if and only if the set is dominated by an integrable random variable; and it is bounded if and only if the set admits an almost sure finite constant upper bound.

With the topological, algebraic, and order structures introduced above, $\CB$, endowed with the topology induced by $d_H$, is a preordered Abelian topological semigroup; see, for instance, Chapter 3 of \citet{Beer1993}. In the definition below, WC stands for ``worst-case''.

\begin{Def}\label{def:risk}
A functional $R\colon\CB\to\R$ is called a \defin{set risk measure} (SRM). It may satisfy the following properties, for all $X,Y\in \CB$, $\alpha \in \R$, $0\le\lambda \le1$, and $\gamma \ge0$:

\begin{enumerate}[label = (\roman*),noitemsep]

\item\label{def:risk monotonicity} \defin{Monotonicity}: If $X \leq Y$, then $R(X) \geq R(Y)$.

\item\label{def:risk trans inv} \defin{Translation invariance}: $R(X+\singleton{\alpha }) = R(X)-\alpha $.

\item\label{def:risk wc bounded} \defin{\wc{}}: $R(X)\leq\sup\nolimits_{x\in X}R(\singleton{x})$.

\item\label{def:risk convex} \defin{Convexity}: $R(\lambda X+(1-\lambda)Y)\leq \lambda R(X)+(1-\lambda)R(Y)$.

\item\label{def:risk ph} \defin{Positive homogeneity}: $R(\gamma X) = \gamma R(X)$.

\end{enumerate}

An SRM is said to be \defin{monetary} if it satisfies \cref{def:risk monotonicity,def:risk trans inv} above. It is called \defin{convex} if it is monetary and satisfies \cref{def:risk convex}, and \defin{coherent} if it is convex and satisfies \cref{def:risk ph}. Throughout the rest of the paper, without loss of generality we assume the following:
\begin{description}
\item \textbf{Normalization}: All SRMs are \defin{normalized}, i.e.\ $R(\singleton{0}) = 0$.
\end{description}
\end{Def}

\begin{Rmk}
Normalization does not affect our main results. It is automatic under positive homogeneity; moreover, if $R$ is monetary, then $R'(X) = R(X)-R(\singleton{0})$ is monetary, normalized, and inherits any other property of $R$ listed in \Cref{def:risk}. The scalar map $\rho(x) = R(\singleton{x})$ is a \emph{bona fide} risk measure, and we call $\rho$ monetary, convex, or coherent according to the corresponding property of $R$.
\end{Rmk}

Monotonicity captures the idea that if, for every random variable $y$ in $Y$ (which may represent a financial position), one can find another random variable $x$ in $X$ that is worse than $y$ (in the sense that it pays less in every possible state of the world), then $X$ is deemed riskier than $Y$. This aligns with the traditional notion of monotonicity for risk measures, where almost surely worse positions are associated with greater risk. This property arises naturally when $X$ models uncertainty (see \Cref{ex: uncertainty sets}) and implies $R(\singleton{0})\leq R(X-X)$ for any $X\in\CB$. Moreover, since $X \subseteq Y$ implies $X \geq Y$, monotone SRMs can be seen as an ``aversion to flexibility'', as in this case $X \subseteq Y$ implies $R(X)\le R(Y)$. In particular,
\begin{equation*}
R(X\cup Y)\geq \max\lbrace R(X),R(Y)\rbrace \geq \min\lbrace R(X),R(Y)\rbrace \geq R(X\intersection Y),
\end{equation*}
where the last inequality makes sense when $X\intersection Y\not = \varnothing$.

As discussed in \Cref{rmk: orders}, monotonicity could be formulated using alternative preorders, such as $\lesssim$. We use the standard preorder $\leq$ because it is the weakest preorder consistent with the principle that worse positions entail higher risk, and because it is compatible with worst-case SRMs (see \Cref{sec:WC}). Using $\lesssim$ would reverse the set-inclusion implication, giving $X \subseteq Y \Rightarrow R(X) \geq R(Y)$; this corresponds to a \textit{preference for flexibility}, as studied in \citet{kreps1979representation}.

Translation invariance is the key axiom behind the interpretation of risk as a capital requirement and behind acceptance-set representations. It says that adding a sure gain to every position in $X$ reduces risk by exactly the same amount; equivalently, $\lambda \mapsto R(X+\singleton{\lambda})$ is strictly decreasing and affine. In some contexts, such as cash-subadditive or systemic risk measures (see \Cref{example systemic risk}), one may want to relax or modify this axiom. We leave such extensions for future work. If $R$ is monetary, then $R$ is finite-valued because $X$ is bounded.

\wc{} means that rendering a set of positions acceptable requires no more capital than the largest amount required by one of its individual positions. This worst-case amount simultaneously collateralizes all positions in the set. The same idea appears in \citet{Fadina2023} under the name ``scenario upper bound''.

Positive homogeneity states that scaling every position in the set by a positive factor $\gamma$ scales the risk of the set by the same factor. Convexity is likewise the direct set-valued analogue of the scalar axiom: it formalizes diversification by requiring that convex combinations of two sets do not increase risk beyond the corresponding convex combination of their risks.

The Hausdorff topology is natural, but too strong for some duality arguments. We therefore also use \emph{Wijsman-type} convergence \citep{Aubin2009,Beer1993}, which records convergence of distance functionals and interacts well with convexification. Let $(X_\alpha)_{\alpha\in A}$ be a net in $\Pow(L^\infty)$. We say that $X_\alpha$ \defin{Wijsman-converges} to $X\subseteq L^\infty$, and write $\Wlim X_\alpha = X$, if $X$ is closed and $\di(\singleton{x},X_\alpha)\to \di(\singleton{x},X)$ for every $x\in L^\infty$. In particular, $\di(\singleton{x},X_\alpha)\to 0$ for every $x\in X$. Hausdorff convergence implies Wijsman convergence of the closures: if $d_H(X_n,X)\to0$, then $\Wlim X_n = \cl(X)$, by Corollary 5.1.11 and Theorem 5.2.10 of \citet{Beer1993}. Hausdorff and Wijsman convergence coincide if and only if the underlying space is totally bounded \citep{Beer1993}; thus they are distinct notions in $L^\infty$. Finally, we write $\Wblim X_\alpha = X$ instead of $\Wlim X_\alpha = X$ when $\{X_\alpha : \alpha \in A\}$ is a bounded subset of $(\CB, d_H)$, and call this \defin{bounded Wijsman convergence}.

\begin{Lmm}\label{lmm: set mono conv}
Let $R$ be a monetary SRM. Then $R$ satisfies

\begin{enumerate}[label = (\roman*),noitemsep]

\item\label{lemma propiedades antigas set-mono} \defin{Set-monotonicity:} If $X,Y\in\CB$ and $X \subseteq Y$, then $R(X) \leq R(Y)$.

\item\label{lemma propiedades antigas lips} \defin{Lipschitz}-property with respect to the Hausdorff distance: $|R(X) - R(Y)| \leq d_H(X,Y)$.
\end{enumerate}

Moreover, if $R$ is convex, then the following equality holds for every compact $X \in \CB$:

\begin{enumerate}[label = (\roman*),noitemsep,resume]

\item\label{lemma propiedades antigas set-con} \defin{Set-convexity:} $R(\conv(X)) = R(X)$.
\end{enumerate}

In addition, if $R$ is lower semicontinuous with respect to bounded Wijsman convergence, then it is set-convex on all of $\CB$.
\end{Lmm}

\begin{proof}
\Cref{lemma propiedades antigas set-mono} follows directly from monotonicity, as $X \subseteq Y$ implies $ X \geq Y$.

For \Cref{lemma propiedades antigas lips}, from the definition of $d_H$ it holds, for all $y \in Y$ and $\epsilon >0$, that $d_H (X,Y)+ \epsilon \geq \inf_{x\in X} \Vert x - y\Vert_\infty $. This implies that for all $y \in Y$ and all $\epsilon >0$ there is some $x \in X$ such that $d_H (X,Y)+ \epsilon \geq \Vert x - y\Vert_\infty \geq x-y$ which, in turn, is equivalent to the following: for all $y \in Y$ and all $\epsilon > 0$ there is some $x \in X$ such that $x \leq y + d_H (X,Y)+ \epsilon $. Hence, $X \leq Y + d_H (X,Y) + \epsilon $. Then, as $R$ is monetary, it holds that $R(X) \geq R(Y + d_H (X,Y) + \epsilon) = R(Y)- d_H (X,Y)-\epsilon $ and $R(Y) - R(X) \leq d_H(X,Y)+ \epsilon $. Reversing the roles of $X$ and $Y$ and taking $ \epsilon $ to $0$ yields the claim.

For \Cref{lemma propiedades antigas set-con}, we shall begin proving the claim for finite sets $F \in \CB$. We first show that, for any such $F$, there is some $\lambda \in (0,1)$ such that $ \lambda F + (1-\lambda) \conv(F) = \conv(F)$. Recall that the operator $\conv$ is the \textit{closed} convex hull, and we are also taking the \textit{closed} sum. Thus, the inclusion $ \lambda F + (1-\lambda) \conv(F) \subseteq \conv(F)$ clearly holds for any $\lambda \in(0,1)$. Now, to show that, for some $\lambda \in(0,1)$, it holds that $ \lambda F + (1-\lambda) \conv(F) \supseteq \conv(F)$, denote by $\alpha $ the number of elements in $F$, take some $\lambda \in (0,\frac{1}{\alpha })$ and $y \in \conv(F)$. Then there is some collection of coefficients $\gamma_i \in [0,1],\, 1 \leq i \leq \alpha $, such that $y = \sum_{i = 1}^{\alpha } \gamma_i x_i,\, x_i \in F $ and $\sum_{i = 1}^{\alpha } \gamma_i = 1$. Clearly, there is at least one $j$ such that $\gamma_j \geq \frac{1}{\alpha }\geq\lambda$, and without loss of generality take $j = 1$. Now let
\begin{align*}
\widehat{\gamma }_1 = \dfrac{\gamma_1 - \lambda}{1-\lambda} \qquad\text{and}\qquad \widehat{\gamma }_i = \dfrac{\gamma_i}{1-\lambda},\quad{i>1}.
\end{align*}
Then the $\widehat{\gamma }_i$ are convex weights and $\hat{y} : = \sum_{i = 1}^{\alpha } \widehat{\gamma }_i x_i \in \conv(F)$. Finally, $y = \lambda x_1 + (1-\lambda) \hat{y}\in \lambda F + (1-\lambda) \conv(F) $. This yields the desired set inclusion.

We have by \cref{lemma propiedades antigas set-mono} that $R(F) \leq R(\conv(F))$. Assume that, for some finite $F$, it holds that $R(F) < R(\conv(F))$. Then the previous reasoning and convexity of the SRM yield, for some $\lambda \in (0,1)$,
\begin{align*}
R(\conv(F)) & = R(\lambda F + (1-\lambda)\conv(F)) \\& \leq \lambda R(F) + (1-\lambda) R(\conv(F)) < R(\conv(F)),
\end{align*}
which is clearly a contradiction. Hence $R(F) = R(\conv(F))$ for finite sets.

As the set of finite sets is dense in the set of compact sets with respect to the Hausdorff distance, and since $R$ is Lipschitz, it follows that, for any compact set $X$, there is a sequence of finite sets $F_n$ such that $X = \Hlim F_n$. Furthermore, the closed convex hull operator is non-expansive with respect to the Hausdorff metric, meaning $d_H(\conv(A), \conv(B)) \leq d_H(A, B)$ for any bounded sets $A, B$ \citep[see][Sections 1.5 and 3.2]{Beer1993}. Consequently, $\Hlim F_n = X$ implies $\Hlim \conv(F_n) = \conv(X)$. Hence, $R(X) = \lim R(F_n) = \lim R(\conv(F_n)) = R(\conv(X)) $.

For the last statement in \Cref{lemma propiedades antigas set-con}, recall that we assume $R$ to be set-monotone (monetary), convex, and l.s.c. with respect to bounded Wijsman convergence. Moreover, we have shown that $R(F) = R(\conv(F))$ holds for any non-empty finite $F$. We shall now extend this property for any infinite set $X\in \CB$.

Fix $X\in \CB$ with infinite cardinality and let $\rvsetset{I}_X$ denote the collection of all finite subsets of $X$. Then $\rvsetset{I}_X$ is a directed set, and $\rvsetset{F}_X: = \big(\conv(F)\big)_{F \in \rvsetset{I}_X}$ is a net indexed by $\rvsetset{I}_X$. Also, as $\cup_{F \in \rvsetset{I}_X} \conv(F)\subseteq\conv(X)$ and $\conv(X)$ is a bounded set, it follows that the net $\rvsetset{F}_X$ is uniformly bounded. Moreover, $\cup_{F \in \rvsetset{I}_X} \conv(F)$ is dense in $\conv(X)$, that is $\cl(\cup_{F \in \rvsetset{I}_X} \conv(F)) = \conv(X)$. Furthermore, since $\rvsetset{I}_X$ is directed by inclusion, whenever $F, G \in \rvsetset{I}_X$ satisfy $F\subseteq G$, then $\conv(F) \subseteq \conv(G) \subseteq \conv(X)$, with $\conv(F), \conv(G) \in \rvsetset{F}_X$. Hence, for every $x\in L^\infty$, $\di(\singleton{x},\conv(G)) \le \di(\singleton{x},\conv(F)),$ whenever $ F\subseteq G$. Therefore, the net $\big(\di(\singleton{x},\conv(F))\big)_{\conv(F)\in \rvsetset{F}_X}$ is decreasing and bounded from below by $0$. Consequently, $ \lim_F \di(\singleton{x},\conv(F)) = \inf_{F\in \rvsetset{I}_X} \di(\singleton{x},\conv(F)),$ and, in addition,
\begin{align*}
\inf_{F\in \rvsetset{I}_X}
\di(\singleton{x},\conv(F))
& = 
\di\left(
\singleton{x},
\bigcup\nolimits_{F\in \rvsetset{I}_X}
\conv(F)
\right)
\\
& = 
\di\left(
\singleton{x},
\cl\left(
\bigcup\nolimits_{F\in \rvsetset{I}_X}
\conv(F)
\right)
\right).
\end{align*}
Therefore, for every $x\in L^\infty$, $ \lim_F \di(\singleton{x},\conv(F)) = \di(\singleton{x},\conv(X)).$ Thus, by the definition of Wijsman convergence, it follows that $\Wlim_F \conv(F) = \conv(X)$. Moreover, due to boundedness, it holds that $\Wblim_F \conv(F) = \conv(X)$. Then lower semicontinuity of $R$ yields the inequality $R(\conv(X)) \leq \liminf_{F \in \rvsetset{I}_X} R(\conv(F))$. Now, since $R(\conv(F)) = R(F)$, we get by set-monotonicity that $R(\conv(F)) = R(F)\leq R(X)$ for any $F \in \rvsetset{I}_X$. Therefore,
\begin{align*}
R(X) \leq R(\conv(X)) \leq \liminf_{F \in \rvsetset{I}_X} R(\conv(F))\leq R(X),
\end{align*}
which yields the desired equality.
\end{proof}

Set-monotonicity and set-convexity, introduced in \Cref{lmm: set mono conv}, are weaker structural properties that play a central role for SRMs. Set-monotonicity is indispensable for characterizing worst-case risk measures. \citet{Fadina2023} call the analogous property ``uncertainty aversion'', with the same interpretation adopted here: greater uncertainty entails greater risk. We stress that set-monotonicity is structural for the present framework. Changing the underlying preorder may break it, in which case the main arguments developed below would no longer apply.

Set-convexity means that diversifying among positions already represented in the set does not change risk. In the theory of preferences over sets of distributions, \citet{dekel2001representing} call the analogous property \textit{Indifference to Randomization}. For us, set-convexity allows an SRM to be identified with a set-function on $\CBC$, which is the key reduction used in the dual representations of \Cref{sec:dual}. The converse question, namely when set-convexity implies convexity, is addressed in \Cref{sec:WC}.

\begin{Rmk}
If $X\in\CB$, then generally $X^{\complement}\notin \CB$. Thus, properties for other set operations such as $R(X^{\complement})$, $R(X\setminus Y)$ and $R(X\triangle Y)$ are outside our proposed scope. While such operations may be relevant---particularly for risk budgeting and risk sharing problems---we leave them for future work. Under the assumptions of \Cref{lmm: set mono conv}, however, $R$ can be extended to bounded, not necessarily closed, sets by defining $R(X) = R(\cl(X))$, since $d_H(X,\cl(X)) = 0$. Other algebraic and order properties of scalar risk measures, such as quasi-convexity, cash-subadditivity, relevance, surplus invariance, and star-shapedness, can be adapted directly. By contrast, distributional properties such as law invariance and comonotonic additivity require a separate discussion of what the distribution of a set of random variables should mean, possibly beyond finite-dimensional distributions. We leave this question for future research.
\end{Rmk}

We now introduce a definition that emulates the traditional concept of acceptance sets. \Cref{prp:acceptance} establishes a link between the properties of an SRM and its acceptance set.

\begin{Def}
The \defin{acceptance set} of an SRM $R$ is defined as
\begin{equation}\label{eq:AR}
\rvsetset{A}_R = \left\lbrace X\in \CB\colon{}R(X)\leq 0 \right\rbrace.
\end{equation}
\end{Def}

\begin{Prp}\label{prp:acceptance}
Let $R$ be a monetary SRM. Then,

\begin{enumerate}[label = (\roman*),noitemsep]

\item\label{prp:acceptance item basic} $\rvsetset{A}_R$ is non-empty and satisfies the following:

\begin{enumerate}

\item if $X\in\rvsetset{A}_R$ and $Y\geq X$, then $Y\in\rvsetset{A}_R$.

\item\label{prp:acceptance item set-monotone...} if $X\in\rvsetset{A}_R$ and $Y\subseteq X$ then $Y\in\rvsetset{A}_R$.

\item $\inf\lbrace \alpha \in\R \colon \singleton{\alpha }\in \rvsetset{A}_R\rbrace = 0$.
\end{enumerate}

\item\label{prp:acceptance item representacao como inf} for all $X\in\CB$,
\begin{equation}\label{eq:risk}
R(X) = \min\lbrace \alpha \in\R \colon X+\singleton{\alpha }\in \rvsetset{A}_R\rbrace.
\end{equation}

\item\label{prp:acceptance item lsc} $R$ is lower semicontinuous with respect to either the Hausdorff metric topology or bounded Wijsman convergence if and only if $\rvsetset{A}_R$ is closed with respect to the corresponding notion of convergence.

\item\label{prp:acceptance item atomic-bounded} $R$ is \Wc{} if and only if, for each $X\in \CB$, the following implication holds: if $\singleton{x}\in \rvsetset{A}_R$ for all $x\in X$, then $X\in \rvsetset{A}_R$.

\item\label{prp:acceptance item convex} $R$ is convex if and only if $\rvsetset{A}_R$ is a convex set in $\CB$.

\item\label{prp:acceptance item positive homogeneous} $R$ is positive homogeneous if and only if $\rvsetset{A}_R$ is a cone in $\CB$.
\end{enumerate}
\end{Prp}

\begin{proof}
For \cref{prp:acceptance item basic}, the claim directly follows from the fact that $R$ is monetary, normalized and that $Y \subseteq X$ implies $Y \geq X$.

For \cref{prp:acceptance item representacao como inf}, note that $X+\singleton{\alpha}\in\CB$ for every $X\in\CB$ and $\alpha\in\R$. A direct calculation gives
\begin{equation*}
\inf\lbrace \alpha \in\R \colon X+\singleton{\alpha }\in \rvsetset{A}_R\rbrace = \inf\lbrace \alpha \in\R \colon R(X)\leq \alpha \rbrace = R(X),
\end{equation*}
for any $X\in\CB$. Since $X+\singleton{R(X)}\in\rvsetset{A}_R$, the infimum is attained.

For \cref{prp:acceptance item lsc}, if $R$ is lower semicontinuous, then $\rvsetset{A}_R = R^{-1}\big((-\infty,0]\big)$ is closed. Conversely, suppose $\rvsetset{A}_R$ is closed. By translation invariance, for every $\alpha\in\R$,
\begin{align*}
&\lbrace X\in\CB\colon R(X)\leq \alpha \rbrace \\ = &\lbrace X\in\CB\colon R(X+\singleton{\alpha })\leq 0\rbrace
 = \lbrace X - \singleton{\alpha }\colon X\in\rvsetset{A}_R\rbrace.
\end{align*}
Thus, all lower sublevel sets of $R$ are closed, which implies that it is lower semicontinuous.

For \cref{prp:acceptance item atomic-bounded}, first let $R$ be \Wc{} and let $X\in \CB$ be such that $\singleton{x}\in\rvsetset{A}_R$ for all $x\in X$. Then $R(X)\leq\sup\nolimits_{x\in X}R(\singleton{x})\leq0$ guarantees that $X\in \rvsetset{A}_R$. For the ``if'' part of the statement, fix $\alpha \in \R$ and $ X\in \CB $ such that $\singleton{x+\alpha }\in\rvsetset{A}_R$ for all $x\in X$. Then $X+\singleton{\alpha } = \bigcup_{x\in X}\singleton{x+\alpha }\in\rvsetset{A}_R$. Thus, we get
\begin{align*}
R(X)& = \inf\lbrace \alpha \in\R \colon X+\singleton{\alpha }\in\rvsetset{A}_R\rbrace \\
&\leq\inf\lbrace \alpha \in\R \colon \singleton{x+\alpha }\in\rvsetset{A}_R, \text{ for all }x\in X\rbrace\\
& = \inf\lbrace \alpha \in\R \colon R(\singleton{x})\leq \alpha, \text{ for all }x\in X\rbrace
\\
& = \sup_{x\in X} R(\singleton{x}).
\end{align*}

The ``only if'' parts of \cref{prp:acceptance item convex,prp:acceptance item positive homogeneous} are immediate. The converse implications follow, respectively, from \cref{prp:acceptance2 item convex,prp:acceptance2 item cone} in \Cref{prp:acceptance2}.
\end{proof}

We now consider the converse problem of generating an SRM from a given collection of acceptable ensembles of positions. First, we define the {induced SRM} of an acceptance family, in analogy with the traditional construction, and then in \Cref{prp:acceptance2} we study how this induced SRM inherits properties of the underlying set.

\begin{Def}
If $\rvsetset{A}\subseteq \CB$ is non-empty, then its \defin{induced SRM} is defined as
\begin{equation}\label{eq:RA}
R_\rvsetset{A}(X) = \inf\lbrace \alpha \in\R \colon X+\singleton{\alpha }\in \rvsetset{A}\rbrace,\quad X\in\CB.
\end{equation}
\end{Def}

\begin{Prp}\label{prp:acceptance2}
Let $\rvsetset{A}\subseteq \CB$ satisfy the following:
\begin{enumerate}[label = \arabic*., noitemsep]

\item\label{prp:acceptance2-item-1} if $X\in\rvsetset{A}$ and $Y\geq X$, then $Y\in\rvsetset{A}$.

\item\label{prp:acceptance2-item-2} $\inf\lbrace \alpha \in\R \colon \singleton{\alpha }\in \rvsetset{A}\rbrace = 0$.
\end{enumerate}
Then:
\begin{enumerate}[label = (\roman*), noitemsep]

\item\label{prp:acceptance2 item monetary} $R_\rvsetset{A}$ is a monetary SRM.

\item\label{prp:acceptance2 item identity if another monetary} if $R$ is an arbitrary monetary SRM, then $R_{\rvsetset{A}_R} = R$.

\item\label{prp:acceptance2 item lsc}
for either the Hausdorff metric topology or bounded Wijsman convergence, $\rvsetset{A}$ is closed if and only if $R_{\rvsetset{A}}$ is lower semicontinuous and $\rvsetset{A}_{R_{\rvsetset{A}}} = \rvsetset{A}$. In either topology, if $R_{\rvsetset{A}}$ is lower semicontinuous, then $\rvsetset{A}_{R_{\rvsetset{A}}} = \cl(\rvsetset{A})$.

\item\label{prp:acceptance2 item atomic-bounded} if, for every $X\in\CB$, the implication holds that $X\in\rvsetset{A}$ whenever $\singleton{x}\in\rvsetset{A}$ for all $x\in X$, then $R_\rvsetset{A}$ is \Wc{}.

\item\label{prp:acceptance2 item convex} if $\rvsetset{A}$ is a convex set, then $R_\rvsetset{A}$ is a convex SRM.

\item\label{prp:acceptance2 item cone} if $\rvsetset{A}$ is a cone, then $R_\rvsetset{A}$ is a positive homogeneous SRM.

\end{enumerate}
Moreover, if $\rvsetset{A}$ is closed with respect to the chosen convergence structure, then the converse implications in \cref{prp:acceptance item atomic-bounded,prp:acceptance item convex,prp:acceptance item positive homogeneous} also hold. In this case, the infimum in \cref{eq:RA} is attained.
\end{Prp}

\begin{proof}

For \Cref{prp:acceptance2 item monetary}, we first show that $R_{\rvsetset{A}}(X)$ is finite for any $X \in \CB$. Since $X$ is bounded, there exists $\alpha > 0$ such that $\singleton{-\alpha } \leq x \leq \singleton{\alpha }$ for all $x \in X$, in particular $\singleton{-\alpha } \leq X \leq \singleton{\alpha }$. Furthermore, $\singleton{\epsilon } \in \rvsetset{A}$ for any $\epsilon > 0$. Since $X + \singleton{\alpha + \epsilon } \geq \singleton{\epsilon }$, it follows that $X + \singleton{\alpha + \epsilon } \in \rvsetset{A}$, yielding $R_{\rvsetset{A}}(X) \leq \alpha + \epsilon < \infty$. Conversely, if $X + \singleton{\beta } \in \rvsetset{A}$, then $\singleton{\beta + \alpha } \geq X + \singleton{\beta }$ implies $\singleton{\beta + \alpha } \in \rvsetset{A}$. We have then that $\alpha + \beta \geq 0$, so $\beta \geq -\alpha $. Taking the infimum yields $R_{\rvsetset{A}}(X) \geq -\alpha > -\infty$. That $R_\rvsetset{A}$ is monetary is straightforward from the properties of $\rvsetset{A}$.

\Cref{prp:acceptance2 item identity if another monetary} follows from \Cref{prp:acceptance}, \cref{prp:acceptance item representacao como inf}.

For \cref{prp:acceptance2 item lsc}, fix either the Hausdorff metric topology or bounded Wijsman convergence. In both cases, scalar translations satisfy $\alpha_i\to\alpha \Rightarrow X + \singleton{\alpha_i} \to X + \singleton{\alpha }$. Throughout this proof, $(\alpha_i)_{i\in I}$ represents a net and ``$\to$" represents either $\Hlim$ or $\Wblim$. Note that since $R_\rvsetset{A}$ is a monetary set risk measure, \Cref{prp:acceptance}, \cref{prp:acceptance item lsc} applies to it.

First assume $\rvsetset{A}$ is closed. Then, by the continuity of scalar translations, $X + \singleton{R_\rvsetset{A}(X)} \in \rvsetset{A}$. Next, we show $\rvsetset{A}_{R_\rvsetset{A}} = \rvsetset{A}$. If $X \in \rvsetset{A}_{R_\rvsetset{A}}$, then $R_\rvsetset{A}(X) \leq 0$, and $X \geq X + \singleton{R_\rvsetset{A}(X)} \in \rvsetset{A}$, which implies $X \in \rvsetset{A}$. Conversely, if $X \in \rvsetset{A}$, then $R_\rvsetset{A}(X) \leq 0$, so $X \in \rvsetset{A}_{R_\rvsetset{A}}$. Thus, $\rvsetset{A}_{R_\rvsetset{A}} = \rvsetset{A}$. Since $\rvsetset{A}$ is closed, $\rvsetset{A}_{R_\rvsetset{A}}$ is closed. By \Cref{prp:acceptance}, \cref{prp:acceptance item lsc}, this implies $R_\rvsetset{A}$ is lower semicontinuous.

Now, assume $R_\rvsetset{A}$ is lower semicontinuous. By \Cref{prp:acceptance}, \cref{prp:acceptance item lsc}, the acceptance set $\rvsetset{A}_{R_\rvsetset{A}}$ is closed. Since it trivially holds that $\rvsetset{A} \subseteq \rvsetset{A}_{R_\rvsetset{A}}$, it follows that $\cl(\rvsetset{A}) \subseteq \rvsetset{A}_{R_\rvsetset{A}}$.

To see the reverse inclusion $\rvsetset{A}_{R_\rvsetset{A}} \subseteq \cl(\rvsetset{A})$, let $X \in \rvsetset{A}_{R_\rvsetset{A}}$, meaning $R_\rvsetset{A}(X) \leq 0$. There is a decreasing net $\alpha_i \to R_\rvsetset{A}(X)$, such that $X + \singleton{\alpha_i} \in \rvsetset{A}$. Let $X_i = X + \singleton{\alpha_i} - \singleton{R_\rvsetset{A}(X)}$. Since $R_\rvsetset{A}(X) \leq 0$, it follows that
\begin{equation*}
X_i = \bigl(X + \singleton{\alpha_i}\bigr) + \singleton{-R_\rvsetset{A}(X)} \geq X + \singleton{\alpha_i}.
\end{equation*}
By the monotonicity of $\rvsetset{A}$, this implies $X_i \in \rvsetset{A}$ for all $i$. By the continuity of scalar translations, $\alpha_i - R_\rvsetset{A}(X) \to 0$ implies $X_i \to X$. Thus, $X \in \cl(\rvsetset{A})$, yielding $\rvsetset{A}_{R_\rvsetset{A}} = \cl(\rvsetset{A})$.

\Cref{prp:acceptance2 item atomic-bounded} follows the same steps as \cref{prp:acceptance item atomic-bounded} in \Cref{prp:acceptance}.

For \cref{prp:acceptance2 item convex} we have, given any $\alpha_1,\alpha_2\in\R$ satisfying $X+\singleton{\alpha_1},Y+\singleton{\alpha_2}\in\rvsetset{A}$, that $\lambda (X+\singleton{\alpha_1})+(1-\lambda)(Y+\singleton{\alpha_2})\in\rvsetset{A}$, for all $\lambda \in[0,1]$. Thus, $R_\rvsetset{A}\left(\lambda X+(1-\lambda)Y\right)\leq \lambda \alpha_1+(1-\lambda)\alpha_2$. Hence, by taking the infimum over those $\alpha_1$ and $\alpha_2$, we get the claim.

For \cref{prp:acceptance2 item cone}, let $X\in\CB$ and $\lambda \geq0$. The case $\lambda = 0$ follows from normalization, so we may assume $\lambda>0$. A direct calculation yields
\begin{align*}
R_\rvsetset{A}(\lambda X)
& = \inf\lbrace \alpha \in\R \colon \lambda X+\singleton{\alpha }\in \rvsetset{A}\rbrace
\\ & = \inf\lbrace \alpha \in\R \colon \lambda \big( X+\lambda^{-1} \singleton{\alpha } \big)\in \rvsetset{A}\rbrace
\\ & = \inf\lbrace \alpha \in\R \colon X+\lambda^{-1}\singleton{\alpha }\in \rvsetset{A}\rbrace
\\ & = \inf\lbrace \lambda \alpha \in\R \colon X+\singleton{\alpha }\in \rvsetset{A}\rbrace
\\ & = \lambda R_\rvsetset{A}(X).
\end{align*}

Finally, the last claim holds directly by \Cref{prp:acceptance} and \cref{prp:acceptance item lsc}.
\end{proof}

Interestingly, we have proved the following fact.

\begin{Crl}
If $\rvsetset{A}\subseteq \CB$ is convex, set-monotone, closed with respect to bounded Wijsman convergence and satisfies the condition $\inf\lbrace \alpha \in\R\colon\, \singleton{\alpha }\in\rvsetset{A}\rbrace = 0$, then the following implication holds:
\begin{equation*}
X\in\rvsetset{A}\Rightarrow \conv(X)\in\rvsetset{A}.
\end{equation*}
\end{Crl}
\begin{proof}
By \Cref{prp:acceptance2}, the acceptance family $\rvsetset{A}$ induces a convex and set-monotone SRM $R_{\rvsetset{A}}$. Since $\rvsetset{A}$ is closed with respect to bounded Wijsman convergence, it follows that $R_{\rvsetset{A}}$ is lower semicontinuous with respect to the same convergence. Hence, \Cref{lmm: set mono conv} implies that $R_{\rvsetset{A}}$ is set-convex. In \cref{lemma propiedades antigas set-con} of \Cref{lmm: set mono conv}, monotonicity in the standard sense is not needed once set-monotonicity holds. Finally, adapting \cref{prp:acceptance2 item lsc} in \Cref{prp:acceptance2} to the bounded Wijsman topology gives $\rvsetset{A}_{R_{\rvsetset{A}}} = \rvsetset{A}$. The claim follows.
\end{proof}

\section{Dual representations}\label{sec:dual}

Throughout this section, the term \emph{measure} is used broadly: it may refer to finitely additive, countably additive, or $\tau$-additive set functions, with the relevant additivity specified in each case. By contrast, a \emph{probability measure} is always understood to be $\sigma$-additive. On each space of finitely additive signed measures introduced below, we use the topology induced by the total variation norm. For a measure $\nu$ on a space $S$, this norm is $\Vert \nu\Vert_{\textsc{tv}} = \sup_{\mathcal{F}} \sum_{S'\in\mathcal{F}}\left\vert\nu(S')\right\vert$, where $\mathcal{F}$ ranges over all finite measurable partitions of $S$. We denote the topological dual $(L^\infty)'$ of $L^\infty\equiv L^\infty(p)$ by $\ba\equiv\ba(\sigmafield{F})$; this is the space of finitely additive signed measures on $(\Omega,\sigmafield{F})$ that are absolutely continuous with respect to $p$ and have finite total variation. The canonical bilinear form on $\ba\times L^\infty$ is written $\E{q}{x} = \int_{\Omega}x\,\dd q$. For any $Q\subseteq\ba$, write $\Vert Q\Vert\coloneqq\sup_{q\in Q}\Vert q\Vert_{\textsc{tv}}$ when no confusion can arise. We denote by $\ba_{+}$ the non-negative elements of $\ba$, and by $\ba_{1,+}$ the elements of $\ba_{+}$ with unit total variation. $P^{\text{ac}}\subseteq\ba_{1,+}$ is the set of \emph{probability measures} $q$ that are absolutely continuous with respect to $p$, with Radon-Nikodym derivative ${\dd q}/{\dd p}$. We also let $U\subseteq\ba$ denote the closed $\Vert\cdot\Vert_{\textsc{tv}}$-unit ball on $\ba$, and $U_{+} = U\intersection\ba_{+}$.

Let $C_b(U)$ be the Banach lattice consisting of all the real-valued functions $f$ defined on $(U,\Vert\cdot\Vert_{\textsc{tv}})$ such that $f$ is continuous and bounded. Let $\bare\coloneqq\bare(\sigmafield{A}_U)$ be the space of regular, signed finitely additive measures of bounded variation on $\sigmafield{A}_U\subseteq \Pow(\ba)$, the \emph{algebra} generated by the open subsets of $U$; see \citet{Aliprantis2006} for the definition of a regularity. Note that an element $\mu\in\bare$ is a measure on a class of subsets of $\ba$, in particular, $\bare\not\subseteq\ba$. We define $\bare_{1,+}$ similarly to $\ba_{1,+}$ and so on. Now, by Theorem 14.10 in \citet{Aliprantis2006}, we can identify $\bare = \left(C_b(U),\Vert\cdot\Vert_{\infty}\right)'$. We let $\Pbare = \{\mu\in\bare_{1,+}\colon\mu(\ba_{1,+}) = 1\}$. Since $\ba_{1,+}$ is closed in $U$ under the total variation topology, one has $\ba_{1,+}\in\sigmafield{A}_U$ and $\ba_{1,+}\in\sigmafield{B}_U$.

\begin{Rmk}
We work with the weak topological pair $(L^{\infty},\ba)$ rather than the more intuitive weak$^*$ pair $(L^{\infty},L^1)$ because the proof of the main results relies on set functional analysis. Most constructions can be adapted to the more general base domain $\CB[L^\lambda(p)]$, $\lambda\in[1,\infty)$, by replacing $\ba$ with $L^\gamma(p)$, where $1/\lambda+1/\gamma = 1$. In that case, duality is given by $f_Y(X) = \big\langle X,Y\big\rangle = \sup_{x\in X,y\in Y}\E{p}{xy}$, with $X\subseteq L^\lambda$ and $Y\subseteq L^\gamma$.
\end{Rmk}

Let $\Rbar \coloneqq \R\cup\lbrace-\infty,\infty\rbrace$. For each $Q\subseteq \ba$ we have that the functional $f_Q\colon\Pow(L^\infty)\to\Rbar$, defined through
\begin{equation*}
f_Q(X) \coloneqq\E{Q}{X} \coloneqq\sup_{x\in X,q\in Q}\E{q}{x}, \qquad X\subseteq L^\infty,
\end{equation*}
is subadditive, positive homogeneous, and satisfies the Cauchy--Schwarz-type inequality $f_Q(X)\leq\lVert X\rVert\cdot\lVert Q\rVert$. Thus, it is finite when both $X$ and $Q$ are bounded, and monotone for any $Q\subseteq\ba_{+}$. Taking closures and convex hulls does not alter the supremum, as $\E{q}{x}$ is continuous and bilinear; thus, for example $f_{\cl(Q)}(\cl(X)) = f_Q(X)$. Furthermore, when $Q = \singleton{q}$ is a singleton, with a slight abuse of notation we write $\E{q}{X}$ in place of $\E{\singleton{q}}{X}$. In this case, $f_{\singleton{q}}$ coincides with the support function of $X$. When $Q\subseteq\ba_{1,+}$, this map is additive for constants as $f_Q(X+\singleton{c}) = f_Q(X)+c$, and if $Q = \singleton{q}$, with $q\in\ba_{1,+}$, i.e., when $Q$ is a singleton, then the map $f_{\singleton{q}}$ is additive as $f_{\singleton{q}}(X+Y) = f_{\singleton{q}}(X)+f_{\singleton{q}}(Y)$. We also have that $d_H(X,Y) = \sup_{q\in U}\left|\E{q}{X}-\E{q}{Y}\right|$ for any $X,Y\in\CBC(L^\infty)$ (see Corollary 3.2.8 in \citet{Beer1993}).

Furthermore, for each $X\in \CB$, define $\widehat{X}\colon U\to\R$ through
\begin{equation}\label{eq:hat-f}
\widehat{X}(q) \coloneqq \sup_{x\in X} \E{q}{x} \equiv \E{q}{X} \equiv f_{\singleton{q}}(X),\qquad q\in U.
\end{equation}
For each $X\in\CB$, we have that $\widehat{X}\in C_b(U)$. In fact, such maps are Lipschitz continuous since for any $q_1,q_2\in U$ we have that $|\widehat{X}(q_1)-\widehat{X}(q_2)|\le \sup_{x\in X}\|x\|_\infty \|\,q_1-q_2\|_{\textsc{tv}}$. Boundedness then follows from that of $U$.

The preceding dual is a space of finitely additive measures. We now refine duality to obtain a representation under the framework of \emph{bona fide} probability measures. Let $\tauare: = \tauare(\sigmafield{B}_U)$ denote the set of regular and $\tau$-additive signed measures with finite total variation, defined on the Borel $\sigma$-field $\sigmafield{B}_U$ on $U$, with $\tau$-additivity meaning that $\mu\!\left( \bigcup_{\alpha } A_\alpha \right) = \lim_{\alpha } \mu(A_\alpha)$ holds for every increasing \emph{net} of open sets $(A_\alpha)$. We define $\tauare_{+}$ and $\tauare_{1,+}$ analogously to $\bare_{+}$ and $\bare_{1,+}$. We equip $C_b(U)$ with the \defin{\strict{} topology}. On norm-bounded subsets of $C_b(U)$, \strict{} convergence is equivalent to uniform convergence on compact subsets of $U$. As seen in \citet{sentilles1972bounded}, the \strict{} topology is locally convex, and the continuous dual of $\left(C_b(U),\strict{}\right)$ is the space $\tauare$. We also define $\Ptau: = \{\mu\in\tauare_{1,+}\colon \mu(\ba_{1,+}) = 1\}$, i.e.\ probability measures in $\tauare$ that are supported on $\ba_{1,+}$. We say that $R$ is \defin{\strict{}-continuous} at $X\in \CBC(L^\infty)$ if, for every net $(X_\alpha)\subset \CBC(L^\infty)$ such that $\widehat{-X_\alpha }\to\widehat{-X}$ in the \strict{} topology, we have $R(X_\alpha)\to R(X)$. We define \defin{\strict{} lower semicontinuity} in a similar fashion for $X\in\CBC(L^\infty)$, and say that $R$ is \defin{\strict{} lower semicontinuous on $\CB$} if, whenever $\widehat{-\conv(X_\alpha)}\to \widehat{-\conv(X)}$ in the \strict{} topology, one has $R(X)\le \liminf_\alpha R(X_\alpha)$. $R$ is said to be \defin{Lipschitz \strict{}-continuous} if there exist a \strict{}-continuous seminorm $\mathcalboon{p}$ on $(C_b(U),\strict{})$ and a constant $L>0$ such that $|R(X)-R(Y)|\le L\,\mathcalboon{p}(\widehat{-X}-\widehat{-Y})$ for every $X,Y\in \CBC(L^\infty)$.

\begin{Thm}\label{thm:dualSRM}
Let $R\colon\CB\to\R$ be a normalized SRM. Then $R$ is convex and \strict{} lower semicontinuous if and only if it admits the representation
\begin{equation}\label{eq:dualSRM3}
R(X) = \sup_{\mu\in \Ptau}\left\lbrace\int_{\ba_{1,+}} \E{q}{-X}\,\dd\mu(q)-R^*(\mu)\right\rbrace,\quad X\in\CB,
\end{equation}
where $R^*\colon\Ptau\to[0,\infty]$ is the non-negative convex functional defined by
\begin{equation}\label{eq:penSRM3}
R^*(\mu) = \sup_{X\in\rvsetset{A}_R}\int_{\ba_{1,+}} \E{q}{-X}\,\dd\mu(q),\qquad \mu\in \Ptau.
\end{equation}
Moreover, $R^*$ is lower semicontinuous for the relative weak$^*$ topology on $\Ptau$. If $R$ is Lipschitz \strict{}-continuous, then the supremum in \eqref{eq:dualSRM3} is attained.
\end{Thm}

\begin{proof}
The ``if'' implication follows immediately from the representation. Conversely, we first prove that $R$ is set-convex. Let $X\in\CB$, and let
$\rvsetset{F}_X$ be the directed family of finite subsets of $X$. For
$F\in\rvsetset{F}_X$, put $G_F: = \conv(F)$. Then
$\widehat{-G_F}\uparrow \widehat{-X}$ pointwise on $U$. Since $X$ is
bounded, the net $\widehat{-G_F}$ is uniformly bounded, and by Dini's theorem
for nets the convergence is uniform on compact subsets of $U$, hence
\strict{}. Therefore,
$R(\conv(X))\leq\liminf_F R(G_F)$. Since $R(G_F) = R(F)$ for
finite $F$, and $F\subseteq X$, set-monotonicity gives
$R(G_F) = R(F)\leq R(X)$. On the other hand, $X\subset\conv(X)$
also gives $R(X)\leq R(\conv(X))$. Hence,
$R(X) = R(\conv(X))$.

Thus it suffices to work on $\CBC(L^\infty)$, the space of non-empty, closed, bounded, and convex sets in $L^\infty$. Addition and multiplication by a non-negative scalar are $d_H$-continuous operations. By \emph{H{\"o}rmander's Theorem} \citep[see Theorem 3.2.9 in][]{Beer1993}, $\CBC(L^\infty)$ is embedded as a closed convex cone $\mathfrak{C}\subseteq C_b(U)$ under the algebraic and isometric embedding $X\mapsto \widehat{X}$. In other words, $(\CBC(L^\infty),d_H)$ is a cone in $(C_b(U),\Vert\cdot\Vert_{\infty})$, with $U$ equipped with the $\Vert\cdot\Vert_{\textsc{tv}}$ topology.

We claim $X\leq Y$ if and only if $\widehat{-X}\geq\widehat{-Y}$ pointwise in $\ba_{1,+}$. We first show it for $U_{+}$. Let $X,Y\in \CBC(L^\infty)$. Then $X\le Y$, equivalently $Y\subset X+L^\infty_{+}$, if and only if $\widehat{-X}(q)\ge\widehat{-Y}(q)$ for every $q\in U_{+}$. Indeed, if $y\in Y$ is the norm limit of $x_\alpha +u_\alpha $, with $x_\alpha \in X$ and $u_\alpha \in L^\infty_{+}$, then, for $q\in U_{+}$, $\E{q}{-y} = \lim_\alpha \E{q}{-x_\alpha -u_\alpha }\le \limsup_\alpha \E{q}{-x_\alpha }\le\widehat{-X}(q)$, and taking the supremum over $y\in Y$ gives $\widehat{-Y}(q)\le\widehat{-X}(q)$. Conversely, if the pointwise inequality holds and some $y_0\in Y$ were outside $X+L^\infty_{+}$, separating $y_0$ from this closed convex set would give a nonzero $\ell\in \ba$ with $\ell(y_0)>\sup_{z\in X+L^\infty_{+}}\ell(z)$. Finiteness forces $\ell(u)\le0$ for all $u\in L^\infty_{+}$, so, after normalization, $q: = -\ell\in U_{+}$, and then $\widehat{-Y}(q)\ge\E{q}{-y_0}>\widehat{-X}(q)$, a contradiction. Since support functions are positive homogeneous, by normalizing we can consider only $q\in U_{+}$ such that $q(\Omega) = 1$. Thus, $X\leq Y$ if and only if $\widehat{-X}_{|\ba_{1,+}}\geq\widehat{-Y}_{|\ba_{1,+}}$. Thus, $X\leq Y$ and $Y\leq X$ if and only if $\widehat{-X}_{|\ba_{1,+}} = \widehat{-Y}_{|\ba_{1,+}}$. By monotonicity of $R$, it follows that such equivalent conditions imply $R(X) = R(Y)$.

We then denote by $f$ and $f^\prime$ elements of $C_b(U)$ and $(C_b(U),\strict{})^\prime$, respectively. Identifying $f^\prime$ with $\mu\in\tauare$, the bilinear form is $\big\langle f,f^\prime\big\rangle = f^\prime(f) = \E{\mu}{f} = \int_U f\,\dd\mu$. 
Let $E = (C_b(U),\strict{})$, $J(X) = \widehat{-X}$, and thus $\mathfrak{C} = J(\CBC(L^\infty))$. The sign adjustment in the bilinear form is used to handle the monotonicity of $R$. Put $V: = \operatorname{span}(\mathfrak{C})$, endowed with the topology induced by $E$. Define $\Pi\colon\mathfrak{C}\to\R$ by $\Pi(J(X)) = R(X)$. This is well-defined, since $J(X) = J(Y)$ on $\ba_{1,+}$ implies $X\le Y$ and $Y\le X$, whence $R(X) = R(Y)$. Thus, $\Pi$ is proper. Moreover, $\Pi$ is convex. Let $f_1 = \widehat{-X_1}$ and $f_2 = \widehat{-X_2}$, where $X_1,X_2\in\CBC(L^\infty)$. Thus, $f = \widehat{-X}$, where $X = \lambda X_1+(1-\lambda)X_2\in \CBC(L^\infty)$. Hence, by convexity of $R$ we get $\Pi(f)\leq \lambda \Pi (f_1)+(1-\lambda)\Pi(f_2)$. $\Pi$ is \strict{} lower semicontinuous by the same property for $R$.

Fix $X\in \CBC(L^\infty)$ and $a<R(X)$. Let
\begin{equation*}
\mathcal{E}: = \{(J(Y),m)\in E\times\R:Y\in \CBC(L^\infty),\ m\ge R(Y)\}.
\end{equation*}
Then $\mathcal{E}$ is convex. Let $\overline{\mathcal{E}}^{\,E\times\R}$ be the closure of $\mathcal{E}$ with respect to the product topology on $E\times\R$. We claim that $(J(X),a)\notin\overline{\mathcal{E}}^{\,E\times\R}$. Indeed, if $(J(Y_\alpha),m_\alpha)\in\mathcal{E}$ converges to $(J(X),a)$, then $J(Y_\alpha)\to J(X)$ in the \strict{} topology and $m_\alpha \to a$. By \strict{} lower semicontinuity, $R(X)\le\liminf_\alpha R(Y_\alpha)\le a$, contradicting $a<R(X)$. Separating $(J(X),a)$ from the closed convex set $\overline{\mathcal{E}}^{\,E\times\R}$, there exist $\ell\in E'$, $\beta \in\R$, and $c\in\R$ such that
\begin{equation*}
\ell(J(X))+\beta a>c\ge \ell(J(Y))+\beta m,\qquad (J(Y),m)\in\mathcal{E}.
\end{equation*}
Since $m$ is arbitrary above $R(Y)$, we have $\beta \le0$. Also $\beta \neq0$, otherwise taking $Y = X$ and $m = R(X)$ gives a contradiction. Hence $\beta <0$. Dividing by $-\beta $, there are $\tilde\ell\in E'$ and $b\in\R$ such that $\tilde\ell(J(Y))+b\le R(Y)$ for all $Y\in \CBC(L^\infty)$, while $\tilde\ell(J(X))+b>a$. Since $E' = \tauare$, there is $\mu\in\tauare$ such that $\tilde\ell(f) = \int_U f\,\dd\mu$. Therefore
\begin{equation*}
a< \int_U\E{q}{-X}\,\dd\mu(q) -\sup_{Y\in \CBC(L^\infty)} \left\{\int_U\E{q}{-Y}\,\dd\mu(q)-R(Y)\right\}.
\end{equation*}
Letting $a\uparrow R(X)$, and using the reverse inequality from the definition of the conjugate, yields
\begin{equation*}
R(X) = \sup_{\mu\in\tauare} \left\{ \int_U\E{q}{-X}\,\dd\mu(q) - \sup_{Y\in\CBC(L^{\infty})} \left( \int_U\E{q}{-Y}\,\dd\mu(q)-R(Y) \right) \right\}.
\end{equation*}
Since $R(X) = R(\conv(X))$, this holds for every $X\in\CB$.

It remains to identify the effective dual domain and to justify why the integral may be taken over $\ba_{1,+}$. Let $J_{\ba_{1,+}}(X): = (-\widehat X)_{|\ba_{1,+}}$. By the order characterization above, $X\le Y$ if and only if $J_{\ba_{1,+}}(X)\ge J_{\ba_{1,+}}(Y)$. Hence $J_{\ba_{1,+}}(X) = J_{\ba_{1,+}}(Y)$ implies $X\le Y$ and $Y\le X$, and monotonicity gives $R(X) = R(Y)$. Thus $R$ factors through the restricted embedding $J_{\ba_{1,+}}$. From this point on, the separation argument is applied to the restricted embedded functional, on the linear span of $J_{\ba_{1,+}}(\CBC(L^\infty))$, endowed with the topology induced by $C_b(\ba_{1,+})$ under the \strict{} topology. The resulting dual variables are regular $\tau$-additive measures on $\ba_{1,+}$, which we identify with elements of $\tauare$ concentrated on $\ba_{1,+}$. Let for now $R^\prime(\mu):= \sup_{Y\in\CBC(L^{\infty})} \left( \int_{\ba_{1,+}}\E{q}{-Y}\,\dd\mu(q)-R(Y) \right)$. Only those $\mu$ such that $R^\prime(\mu)<\infty$ matters for the supremum representation. Under this restricted duality  monotonicity of $R$ implies that only positive dual variables have finite conjugate. Indeed, if a dual variable were not positive, the conjugate of the increasing restricted functional would be infinite. Moreover, since $q(\Omega) = 1$ for every $q\in\ba_{1,+}$, translation invariance gives $J_{\ba_{1,+}}(X+\singleton{a}) = J_{\ba_{1,+}}(X)-a$. Hence, for a dual variable $\mu$, the conjugate applied to $X+\singleton{a}$ contains the term $a(1-\mu(\ba_{1,+}))$. Since $a\in\R$ is arbitrary, finiteness of the conjugate forces $\mu(\ba_{1,+}) = 1$. Therefore the effective dual variables are precisely the elements of $\Ptau$, and the linear term is $\int_{\ba_{1,+}}\E{q}{-X}\,\dd\mu(q)$.

For $\mu\in\Ptau$, the acceptance-set penalty formula is now justified. Indeed, by translation invariance, every $X\in\CB$ satisfies $X+\singleton{R(X)}\in\rvsetset{A}_R$, and the linear term shifts by exactly $-R(X)$. Therefore $R^*(\mu) = \sup_{X\in\rvsetset{A}_R} \int_{\ba_{1,+}}\E{q}{-X}\,\dd\mu(q)$. Convexity is straightforward, and non-negativity follows from $\singleton{0}\in\rvsetset{A}_R$. Finally, $R^*$ is weak$^*$ lower semicontinuous on $\Ptau$, endowed with the relative weak$^*$ topology, because it is the pointwise supremum of the weak$^*$-continuous affine maps $\mu\mapsto\int_{\ba_{1,+}}\E{q}{-X}\,\dd\mu(q)$.

It remains to prove attainment of the supremum. Recall that $\Pi$ is defined by $\Pi(J(X)): = R(X)$ on $J(\CBC(L^\infty))$. By \strict{} Lipschitz continuity, choose a \strict{}-continuous seminorm $\mathcalboon{p}$ and $L>0$ such that $|\Pi(J(X))-\Pi(J(Y))|\le L\,\mathcalboon{p}(J(X)-J(Y))$. Define on $(C_b(U),\strict{})$ the map
\begin{equation*}
\widetilde\Pi(f): = \inf\limits_{Y\in\CBC(L^\infty)} \{R(Y)+L\,\mathcalboon{p}(f-J(Y))\}.
\end{equation*}
This infimum is finite. Indeed, fixing any $Y_0\in\CBC(L^\infty)$, the Lipschitz estimate and the triangle inequality give
\begin{equation*}
R(Y)+L\,\mathcalboon{p}(f-J(Y))\ge R(Y_0)-L\,\mathcalboon{p}(f-J(Y_0))
\end{equation*}
for every $Y\in\CBC(L^\infty)$, while the opposite finiteness follows by testing the infimum at $Y_0$. Thus $\widetilde\Pi$ is finite-valued. It is also convex and $L\mathcalboon{p}$-Lipschitz, hence \strict{}-continuous. Moreover, $\widetilde\Pi(J(X)) = R(X)$ for every $X\in \CBC(L^\infty)$: the inequality $\le$ follows by definition, while the inequality $\ge$ follows from the Lipschitz estimate. Since $\widetilde\Pi$ is a finite \strict{}-continuous convex functional on the locally convex space $(C_b(U),\strict{})$, it has a continuous subgradient at every point. Hence, for fixed $X\in \CBC(L^\infty)$, there exists $\mu_X\in\tauare$ such that $\mu_X\in\partial\widetilde\Pi(J(X))$. Thus $\widetilde\Pi(J(X))+\widetilde\Pi^*(\mu_X) = \int_U J(X)\,\dd\mu_X$. Let
\begin{equation*}
\Pi^*(\mu): = \sup\limits_{Y\in\CBC(L^\infty)} \left\{\int_U J(Y)\,\dd\mu-R(Y)\right\}.
\end{equation*}
Since $\widetilde\Pi = \Pi$ on the embedded domain, $\Pi^*(\mu_X)\le\widetilde\Pi^*(\mu_X)<\infty$. Hence the effective-domain identification already proved above applies, and $\mu_X\in\Ptau$. For such $\mu_X$, $\Pi^*(\mu_X) = R^*(\mu_X)$. Moreover, $\int_U J(X)\,\dd\mu_X-\Pi^*(\mu_X)\ge R(X)$, while the reverse inequality follows from the definition of $\Pi^*$, by testing at $Y = X$. Therefore $R(X) = \int_{\ba_{1,+}}\E{q}{-X}\,\dd\mu_X(q)-R^*(\mu_X)$, so the supremum in \eqref{eq:dualSRM3} is attained.
\end{proof}

\begin{Rmk}
For SRMs, \strict{} convergence plays a similar role as Fatou continuity in the scalar theory (equivalent to continuity from above for monetary risk measures), i.e., lower semicontinuity with respect to dominated almost sure convergence, for scalar risk measures. \Cref{fig:horta} exposes a visual structure. For a convex SRM the dual representation holds if and only if $\rvsetset{A}_R$ is closed with respect to the \strict{} convergence. Since in this paper the property holds for nets, it has indeed a stronger consequence. While Fatou continuity is necessary for a dual representation under probabilities for scalar convex risk measures on $C_b$ spaces, it is, in general, not sufficient. For instance, see \citet{GaoLeungXanthos2019} for a discussion on Orlicz spaces. Nonetheless, when $\ba_{1,+}$ is Polish, \citet{Delbaen2022} and \citet{Nendel2024} show that, for traditional risk measures and under additional technical conditions, Fatou continuity is equivalent to a supremum representation over probabilities on the Borel sigma-algebra of $\ba_{1,+}$.
\end{Rmk}

\begin{figure}[htbp]\label{fig:horta}
 \centering
 \begin{tikzpicture}[scale = 0.8, transform shape,
 > = Stealth,
 box/.style = {
 rectangle, draw = MidnightBlue!80, fill = MidnightBlue!4, thick,
 rounded corners = 2mm, minimum height = 1.4cm, minimum width = 4cm,
 align = center, font = \sffamily,
 drop shadow = {opacity = 0.1, shadow xshift = 0.5ex, shadow yshift = -0.5ex}
 },
 arrow_label/.style = {
 font = \footnotesize\sffamily, text = black,
 align = center, inner sep = 6pt
 }
 ]
 \node[box] (funcs) at (0, 3.5) {\textbf{Function Space} \\ $(C_b(U), \textsf{strict})$};
 \node[box] (sets) at (-5.5, 0) {\textbf{Set Space} \\ $\CBC(L^\infty)$};
 \node[box] (measures) at (5.5, 0) {\textbf{Dual Measure Space} \\ $\Ptau \subset \tauare$};
 \node[box] (risk) at (0, -3.5) {\textbf{Risk Value} \\ $\R$};
 
 \draw[->, thick, black] (sets) -- node[above left, arrow_label] {H{\"o}rmander Embedding\\$X \mapsto \widehat{-X}$} (funcs);
 \draw[<->, thick, black] (funcs) -- node[above right, arrow_label] {Fenchel--Moreau\\Duality\\$\Pi \longleftrightarrow \Pi^\prime$} (measures);
 \draw[->, thick, black] (sets) -- node[below left, arrow_label] {Set Risk Measure\\$R(X)$} (risk);
 \draw[->, thick, dashed, black] (measures) -- node[below right, arrow_label] {Dual Representation\\$\sup_{\mu\in\Ptau}\{\int_{\ba_{1,+}} \widehat{-X}(q)\,\dd\mu(q)-R^*(\mu)\}$} (risk);
 \end{tikzpicture}
 \caption{Structural overview of the dual representation framework. The nonlinear hyperspace of sets is embedded into a space of bounded continuous functions, where Fenchel--Moreau duality yields a representation over regular $\tau$-additive measures.}
 \label{fig:duality_diagram}
\end{figure}

\begin{Crl}\label{crl:dualSRM}
An SRM $R$ is coherent and \strict{} lower semicontinuous if and only if
\begin{equation}\label{eq:dualSRMcoh}
R(X) = \sup_{\mu\in \Ptau_R}\int_{\ba_{1,+}} \E{q}{-X}\,\dd\mu(q),\qquad X\in\CB,
\end{equation}
where $\Ptau_R$ is non-empty, weak$^*$ closed, and convex, and is given by
\begin{align*}
\Ptau_R& = \left\lbrace \mu\in \Ptau \colon\int_{\ba_{1,+}}
\E{q}{-X}
\,\dd\mu(q)\leq R(X)\text{ for all }X\in\CB\right\rbrace \\
& = \lbrace \mu\in \Ptau \colon R^*(\mu) = 0\rbrace.
\end{align*}
\end{Crl}
\begin{proof}
From \Cref{thm:dualSRM}, positive homogeneity for $R$ implies that $R^*$ can take values only in $\lbrace 0,\infty\rbrace $. Finiteness of $R$ then ensures the claim and the fact that $\Ptau_R\neq\emptyset$. Convexity is immediate. The set $\Ptau$ is weak$^*$ closed. Indeed, positivity and normalization are weak$^*$ closed. If a weak$^*$-limit $\mu$ of elements of $\Ptau$ charged $U\setminus\ba_{1,+}$, regularity and the closedness of $\ba_{1,+}$ would give $f\in C_b(U)$ with $f = 0$ on $\ba_{1,+}$ and $\int f\,\dd\mu>0$, contradicting $\int f\,\dd\mu = \lim_\alpha \int f\,\dd\mu_\alpha = 0$. Thus $\mu(\ba_{1,+}) = 1$, and $\mu\in\Ptau$. Weak$^*$ closedness follows because $\Ptau_R = \Ptau\cap\{\mu:R^*(\mu) = 0\}$, $\Ptau$ is weak$^*$-closed, and $R^*$ is weak$^*$-lower semicontinuous. Non-emptiness follows from the dual representation of the finite coherent SRM $R$.
\end{proof}

\begin{Crl}\label{crl:wblim}
If $R$ is a convex SRM and is \strict{} lower semicontinuous, then $R$ is lower semicontinuous with respect to bounded Wijsman convergence.
\end{Crl}
\begin{proof}
Let $X = \Wblim_\alpha X_\alpha $, where $X\in\CB$ and $\{X_\alpha \}\subset\CB$ is bounded. By \Cref{thm:dualSRM}, $R$ admits the representation \eqref{eq:dualSRM3}. Fix $\mu\in\Ptau$. We first show that $X\mapsto\int_{\ba_{1,+}}\E{q}{-X}\,\dd\mu(q)$ is lower semicontinuous with respect to bounded Wijsman convergence. Since support functions are not affected by closed convex hulls, we may replace each $X_\alpha $ by $\conv(X_\alpha)$ inside the integral. Fix a finite set $F = \{x_1,\ldots,x_n\}\subset X$ and put $G_F: = \conv(F)$. Since $\di(\singleton{x_i},X_\alpha)\to0$, we may choose $x_{i,\alpha }\in X_\alpha $ with $x_{i,\alpha }\to x_i$ in $\|\cdot\|_\infty$. Let $F_\alpha : = \{x_{1,\alpha },\ldots,x_{n,\alpha }\}$ and $G_\alpha : = \conv(F_\alpha)$. Then $d_H(G_\alpha,G_F)\to0$, and therefore $\widehat{-G_\alpha }\to\widehat{-G_F}$ uniformly on $U$. Since $\mu$ has finite total variation,
\begin{equation*}
\int_{\ba_{1,+}}\E{q}{-G_\alpha }\,\dd\mu(q) \to \int_{\ba_{1,+}}\E{q}{-G_F}\,\dd\mu(q).
\end{equation*}
Moreover, $G_\alpha \subset\conv(X_\alpha)$, so $\E{q}{-G_\alpha }\le \E{q}{-\conv(X_\alpha)} = \E{q}{-X_\alpha }$ for every $q\in\ba_{1,+}$. Hence
\begin{equation*}
\liminf_\alpha \int_{\ba_{1,+}}\E{q}{-X_\alpha }\,\dd\mu(q) \ge \int_{\ba_{1,+}}\E{q}{-G_F}\,\dd\mu(q).
\end{equation*}
It remains to let $F$ increase to $X$. Let $\rvsetset{F}(X)$ be the directed set of finite subsets of $X$, ordered by inclusion. Then $\widehat{-G_F}\uparrow\widehat{-X}$ pointwise on $U$. Since $X$ is bounded, this net is norm-bounded; by Dini's theorem for nets, the convergence is uniform on compact subsets of $U$, hence \strict{}. Since $\mu\in\Ptau\subset\tauare = (C_b(U), \strict{})'$, we get
\begin{equation*}
\int_{\ba_{1,+}}\E{q}{-G_F}\,\dd\mu(q) \to \int_{\ba_{1,+}}\E{q}{-X}\,\dd\mu(q).
\end{equation*}
Combining the previous inequalities yields
\begin{equation*}
\int_{\ba_{1,+}}\E{q}{-X}\,\dd\mu(q) \le \liminf_\alpha \int_{\ba_{1,+}}\E{q}{-X_\alpha }\,\dd\mu(q).
\end{equation*}
Finally, by the representation \eqref{eq:dualSRM3},
\begin{equation*}
R(X) \le \sup_{\mu\in\Ptau} \left\{ \liminf_\alpha \int_{\ba_{1,+}}\E{q}{-X_\alpha }\,\dd\mu(q) -R^*(\mu) \right\} \le \liminf_\alpha R(X_\alpha).
\end{equation*}
Therefore $R$ is lower semicontinuous with respect to bounded Wijsman convergence.
\end{proof}

\section{Worst-case SRMs}\label{sec:WC}

In this section we focus on \emph{worst-case} SRMs, namely those satisfying $R(X) = \sup_{x\in X}R(\singleton{x})$ for all $X\in\CB$.

\begin{Prp}\label{prp:SRM}
Let $R$ be a monetary SRM. Then $R$ is \Wc{} if and only if the \defin{worst-case representation}
\begin{equation}\label{eq:sup}
R(X) = \sup_{x\in X} R(\singleton{x})
\end{equation}
holds for all $X\in \CB$. In this case:
\begin{enumerate}[noitemsep]

\item The acceptance set of $R$ is given by the condition $X\in\rvsetset{A}_R$ if and only if $R(\singleton{x})\le0$ for all $x\in X$, i.e.,
\begin{equation*}
\rvsetset{A}_R = \lbrace X\in \CB \colon R(\singleton{x})\le0\text{ for all }x\in X\rbrace.
\end{equation*}

\item $R$ is lower semicontinuous with respect to bounded Wijsman convergence.
\end{enumerate}
\end{Prp}

\begin{proof}
The representation follows from \Cref{lmm: set mono conv}, since set-monotonicity together with \wc{} yields the formula. The description of $\rvsetset{A}_R$ is immediate. To prove the claim on lower semicontinuity, let $\rho(z) = R(\singleton{z})$. Since $\rho$ is monetary, it is $1$-Lipschitz in $\|\cdot\|_\infty$. Fix $\varepsilon>0$ and choose $z_0\in X$ such that $\rho(z_0)\ge \sup_{z\in X}\rho(z)-\varepsilon$. By Wijsman convergence, $\di(\singleton{z_0},X_\alpha)\to0$. Choose $z_\alpha \in X_\alpha $ with $\|z_\alpha -z_0\|_\infty\to0$. Then
\begin{equation*}
\liminf_\alpha R(X_\alpha) = \liminf_\alpha \sup_{y\in X_\alpha }\rho(y) \ge \liminf_\alpha \rho(z_\alpha) = \rho(z_0) \ge R(X)-\varepsilon.
\end{equation*}
Letting $\varepsilon\downarrow0$ yields the claim.
\end{proof}

\begin{Rmk}
Since SRMs are finite-valued, \wc{} is equivalent to the following implication: if, for each $X\in\CB$, there is a real scalar $\alpha_X$ such that $\sup_{x\in X}R(\singleton{x})\leq \alpha_X$, then $R(X)\leq \alpha_X$. Both properties are implied by the stronger condition of \defin{element monotonicity}: for all $X,Y\in\CB$, if $R(\singleton{x})\leq R(\singleton{y})$ for all $x\in X$ and all $y\in Y$, then $R(X)\leq R(Y)$. That this is indeed stronger follows by taking $Y = \singleton{-\sup_{x\in X}R(\singleton{x})}$. In addition, under monotonicity, \Cref{prp:SRM} implies that all these properties are equivalent, since the worst-case representation \eqref{eq:sup} implies element monotonicity.
\end{Rmk}

\begin{Rmk}\label{Rmk:link between R and rho in the WC setting}
In view of \Cref{prp:SRM}, under the worst-case representation there is a link between SRMs and traditional risk measures on $L^\infty$. Indeed, let $\rho$ be a monetary risk measure on $L^\infty$, and suppose $R$ is defined on $\CB$ through $R(X) = \sup_{x\in X}\rho(x)$. By recalling that the acceptance set of $\rho$ is $A_\rho = \lbrace x\in L^\infty\colon\, \rho(x)\le0\rbrace$, we have by \Cref{prp:SRM} that $\rvsetset{A}_R = \lbrace X\in\CB\colon X\subseteq A_\rho\rbrace $, that is, $\rvsetset{A}_R = \Pow\left(A_\rho\right)\intersection\CB$. Equivalently, for $X\in\CB$ one has $X\in\rvsetset{A}_R$ if and only if $X\subseteq A_\rho$.
\end{Rmk}

\begin{Thm}\label{thm:SRM2}
Let $R$ be a \Wc{} monetary SRM. Then, the following are equivalent:
\begin{enumerate}[label = (\roman*),noitemsep]

\item\label{theorem SRM2: R convex} $R$ is convex.

\item\label{theorem SRM2: rho convex} $x\mapsto R(\singleton{x})$ is convex.

\item\label{theorem SRM2: set-convex} $R$ is set-convex.

\item\label{theorem SRM2: representation} for every $X\in \CB$ one has
\begin{equation}\label{eq:dual2}
R(X) = \sup_{q\in \ba_{1,+}}\left\lbrace \E{q}{-X} -R^*(\delta_q) \right\rbrace,
\end{equation}
where $\delta_q$ denotes the Dirac measure concentrated at $q\in\ba_{1,+}$ and
\begin{equation*}
R^*(\delta_q): = \sup_{X\in\rvsetset{A}_R}\E{q}{-X}.
\end{equation*}
\end{enumerate}
In this case, $R$ is \strict{} lower semicontinuous.
\end{Thm}

\begin{proof}
The equivalence between \ref{theorem SRM2: R convex} and \ref{theorem SRM2: rho convex} is straightforward. \ref{theorem SRM2: R convex} $\Rightarrow$ \ref{theorem SRM2: set-convex} is direct from \Cref{prp:SRM} and \Cref{lmm: set mono conv}. For \ref{theorem SRM2: set-convex} $\Rightarrow$ \ref{theorem SRM2: rho convex}, let $x,y\in L^\infty$ and $\lambda \in[0,1]$. Then,
\begin{equation*}
R(\singleton{\lambda x+(1-\lambda)y})\leq R(\conv(\lbrace x,y\rbrace))\leq R(\lbrace x,y\rbrace) = \max\lbrace R(\singleton{x}),R(\singleton{y})\rbrace.
\end{equation*}
Here set-monotonicity follows from \Cref{lmm: set mono conv}. Thus, $x\mapsto R(\singleton{x})$ is quasi-convex. By translation invariance, it is a convex risk measure in $L^\infty$.

Now, assume \ref{theorem SRM2: representation} holds. Then it is clear that \eqref{eq:dual2} defines a \Wc{}, convex and set-convex SRM. Thus, the equivalences in \Cref{theorem SRM2: R convex,theorem SRM2: rho convex,theorem SRM2: set-convex} hold. For the converse implication, the map $\rho(x): = R(\singleton{x})$ is a convex risk measure in $L^\infty$, with acceptance set $A_\rho = \rho^{-1}(-\infty,0]$. The corresponding \defin{penalty term} is the mapping $\rho^* \colon \ba_{1,+}\to\R_{+}\cup\lbrace\infty\rbrace $, defined through $\rho^*(q) = \sup_{x\in A_\rho}\E{q}{-x}$. Chapter 4 of \citet{Follmer2016} states that $\rho$ can be represented as
\begin{equation*}
\rho(x) = \max_{q\in\ba_{1,+}}\left\lbrace \E{q}{-x}-\rho^*(q) \right\rbrace,\quad x\in L^\infty.
\end{equation*}
We then get that $R(X) = \sup_{x\in X}R(\singleton{x})$ holds if and only if
\begin{align*}
R(X)& = \sup_{x\in X}\sup_{q\in\ba_{1,+}}\left\lbrace \E{q}{-x}-\rho^*(q)\right\rbrace\\
& = \sup_{q\in\ba_{1,+}}\left\lbrace\sup_{x\in X} \E{q}{-x}-\rho^*(q)\right\rbrace\\
& = \sup_{q\in\ba_{1,+}}\left\lbrace\E{q}{-X}-\rho^*(q)\right\rbrace.
\end{align*}
It remains to show that $R$ is \strict{} lower semicontinuous. If $\widehat{-X_i}\to\widehat{-X}$ in the \strict{} topology, we have $\widehat{-X}(q)\leq\liminf_i\widehat{-X_i}(q)$ for any $q\in\ba_{1,+}$. Thus,
\begin{equation*}
R(X)\leq\liminf\limits_i\sup_{q\in \ba_{1,+}}\left\lbrace \E{q}{-X_i} -R^*(\delta_q) \right\rbrace = \liminf\limits_i R(X_i).
\end{equation*}

By \Cref{thm:dualSRM}, every $X\in\CB$---in particular every singleton $\singleton{x}$---has a dual representation in terms of $R^*$. From \Cref{prp:SRM}, for $X\in\CB$, one has $X\in\rvsetset{A}_R$ if and only if $\singleton{x}\in\rvsetset{A}_R$ for all $x\in X$, equivalently $x\in A_\rho$ for all $x\in X$. Thus the penalty term is
\begin{equation*}
\rho^*(q) = \sup_{x\in A_\rho}\E{q}{-x} = \sup_{X\in\rvsetset{A}_R}\sup_{x\in X}\E{q}{-x} = \sup_{X\in\rvsetset{A}_R}\E{q}{-X} = R^*(\delta_q).
\end{equation*}
This concludes the proof.
\end{proof}

\begin{Rmk}
Under \wc{}, \cref{eq:dual2} is obtained from \cref{eq:dualSRM3} by restricting the supremum over $\lbrace \mu\in \Ptau\colon\mu = \delta_q \text{ for some } q\in \ba_{1,+}\rbrace $. Moreover, for convex SRMs satisfying \wc{}, lower semicontinuity with respect to bounded Wijsman convergence is equivalent to \strict{} lower semicontinuity.
\end{Rmk}

\begin{Rmk}
A question of interest here is whether the supremum in the worst-case formulation is attained. Under monotonicity, we have, for any $X\in\CB$ and letting $y_X = \essinf X$, that $\sup_{x\in X}R(\singleton{x})\leq R(\singleton{y_X})$. In particular, when $y_X \in X$, then it holds that $R(X) = \max_{x\in X}R(\singleton{x}) = R(\singleton{y_X})$. Alternatively, if $x\mapsto R(\singleton{x})$ is convex, then we have that
\begin{equation*}
R(X) = R(\conv(X)) = \sup_{x\in \conv(X)} R(\singleton{x}),\qquad X\in \CB.
\end{equation*}
Hence the supremum is attained whenever $\conv(X)$ is compact in a topology under which $x\mapsto R(\singleton{x})$ is upper semicontinuous. For instance, if one works with the weak$^*$ topology $\sigma(L^\infty,L^1)$, with $x\mapsto R(\singleton{x})$ is upper semi-continuous in this topology, under then norm-boundedness of $\conv(X)$ and $\sigma(L^\infty,L^1)$-closedness imply compactness by Banach--Alaoglu.
\end{Rmk}

\begin{Crl}
Let $R$ be a \Wc{} monetary SRM. Then $R$ is coherent if and only if $x\mapsto R(\singleton{x})$ is coherent, and this is equivalent to the representation
\begin{equation}\label{eq:cohdual2}
R(X) = \sup_{q\in \ba_R} \E{q}{-X},\qquad X\in \CB,
\end{equation}
where $\ba_R$ is non-empty, weak$^*$ closed, and convex, and is given by
\begin{align*}
\ba_R & = \lbrace q\in\ba_{1,+}\colon \E{q}{-X}\leq R(X)\,\text{ for all }\,X\in \CB\rbrace \\
& = \lbrace q\in\ba_{1,+}\colon R^*(\delta_q) = 0\rbrace.
\end{align*}
\end{Crl}
\begin{proof}
The equivalence for positive homogeneity is clear. From \eqref{eq:dual2} we get that $R^*$ only takes $0$ and $\infty$ as possible values, and then
\begin{align*}
\ba_R & = \left\lbrace q\in \ba_{1,+}\colon R^*(\delta_q) = 0\right\rbrace\\
& = \lbrace q\in \ba_{1,+}\colon \E{q}{-X}\leq R(X)\text{ for all }X\in \CB\rbrace.
\end{align*}
Non-emptiness of $\ba_R$ follows from the coherent dual representation of the scalar risk measure $\rho(x): = R(\singleton{x})$. Moreover, convexity and weak$^*$ lower semicontinuity of $q\mapsto \E{q}{-X}$ imply that $\ba_R$ is weak$^*$ closed and convex.
\end{proof}

\section{Examples and applications}\label{sec:examp}

\begin{Exm}[Entropic risk measure]
The \defin{entropic risk measure} is a convex risk measure generated by exponential utility on $L^\infty$. It is defined, for a scalar parameter $\gamma >0$, as
\begin{equation}
\label{eq:entropic-risk-measure} \rho_{\gamma }(x) = \frac{1}{\gamma } \log \E{p}{\mathrm{e}^{-\gamma x}},\quad x\in L^\infty.
\end{equation}
Recall that $P^{\text{ac}}\subseteq\ba_{1,+}$ is the set of \emph{probability measures} $q$ that are absolutely continuous with respect to $p$. Its penalty is the relative entropy, defined as
\begin{equation*}
\rho_{\gamma }^*(q) = \frac{1}{\gamma }\E{q}{\log\left(\frac{\dd q}{\dd p}\right)}, \quad q\in P^{\text{ac}}.
\end{equation*}
The supremum in $\rho_{\gamma }(x) = \sup_{q\in P^{\text{ac}}} \big\lbrace\E{q}{-x}-\rho^*_{\gamma }(q)\big\rbrace$ is attained, for $x\in L^\infty$, at $\nicefrac{\dd q_x}{\dd p} = \nicefrac{\mathrm{e}^{-\gamma x}}{\E{p}{\mathrm{e}^{-\gamma x}}}$. We now illustrate how to adapt the concept of entropic risk measures to the framework of SRMs, without resorting to the worst-case setup. To do that, fix some $\mu_0\in\Ptau$ and take $\gamma $ as above, and then define an SRM on $\CB$ as
\begin{equation}
\label{eq:entropic-SRM} R_{\gamma }(X) = \frac{1}{\gamma }\log\int_{\ba_{1,+}}\mathrm{e}^{\gamma \widehat{-X}}\,\dd\mu_0(q).
\end{equation}
It fails \wc{}, as the following two-state example illustrates. Let $\Omega = \{1,2\}$, and let $q_1$ and
$q_2$ be the Dirac probabilities on states $1$ and $2$. Take
$\mu_0 = \frac12\delta_{q_1}+\frac12\delta_{q_2}$. Fix $a>0$ and define
$x_1(1) = -a$, $x_1(2) = 0$, $x_2(1) = 0$, and $x_2(2) = -a$. Let
$X = \{x_1,x_2\}$. Then $\widehat{-X}(q_1) = \widehat{-X}(q_2) = a$, and therefore
$R_{\gamma }(X) = a$. On the other hand,
$\widehat{-\singleton{x_1}}(q_1) = a$, $\widehat{-\singleton{x_1}}(q_2) = 0$, and symmetrically
$\widehat{-\singleton{x_2}}(q_1) = 0$, $\widehat{-\singleton{x_2}}(q_2) = a$. Hence
\begin{equation*}
R_{\gamma }(\singleton{x_1}) = R_{\gamma }(\singleton{x_2}) = \frac1{\gamma }\log\left(\frac{\mathrm{e}^{\gamma a}+1}{2}\right) <a.
\end{equation*}
Hence $R_{\gamma }(X)>\sup_{x\in X}R_{\gamma }(\singleton{x})$, and \wc{} fails.

Using the support-function and integral properties, it is straightforward to verify that $R_{\gamma }$ is convex and lower semicontinuous with respect to both bounded Wijsman convergence and the \strict{} topology. Hence, by \Cref{thm:dualSRM}, it admits the representation
\begin{equation*}
R_\gamma (X) = \sup_{\mu\in \Ptau}\left\lbrace\int_{\ba_{1,+}} \E{q}{-X}\,\dd\mu(q)-R^*_\gamma (\mu)\right\rbrace,
\end{equation*}
where one admissible penalty is ${R^*_{\gamma }}(\mu) = \frac{1}{\gamma }\E{\mu_0}{\frac{\dd\mu}{\dd\mu_0}\log\frac{\dd\mu}{\dd\mu_0}}$ for $\mu\!\ll \mu_0$, and ${R^*_{\gamma }}(\mu) = \infty$ otherwise. By an application of the usual Lagrangian method over the dual pair $(C_b(U),\tauare)$ to the map
\begin{align*}
&\mu\mapsto \E{\mu_0}{\widehat{-X}\frac{\dd\mu}{\dd\mu_0}}-\frac{1}{\gamma }\E{\mu_0}{\frac{\dd\mu}{\dd\mu_0}\log\frac{\dd\mu}{\dd\mu_0}},
\\&\text{s.t.}\:\begin{cases}\frac{\dd\mu}{\dd\mu_0}\geq 0,\\
\E{\mu_0}{\frac{\dd\mu}{\dd\mu_0}} = 1,
\end{cases}
\end{align*}
we have that the maximum is attained at $\mu^*$ with $\frac{\dd\mu^*}{\dd\mu_0} = {\mathrm{e}^{\gamma \widehat{-X}}}\Big/{\E{\mu_0}{\mathrm{e}^{\gamma \widehat{-X}}}}$.
\end{Exm}

\begin{Exm}[Expected Shortfall]
A canonical coherent risk measure on $L^\infty$ is the \defin{level-$\alpha $ Expected Shortfall, denoted $\ES^{\alpha }$} and defined, for $0<\alpha \le1$, as
\begin{equation}
\label{eq:expected-shortfall} \ES^{\alpha }(x) = -\frac{1}{\alpha }\int_0^{\alpha } F^{-1}_x(u)\,\dd u.
\end{equation}
The dual set of $\ES^{\alpha }$ is given by
\begin{equation*}
Q_{\ES^{\alpha }} = \left\lbrace q\in P^{\text{ac}}\colon\frac{\dd q}{\dd p}\leq\frac{1}{\alpha }\right\rbrace.
\end{equation*}
As in the entropic example, fix $\mu_0\in\Ptau$ and define the \defin{Expected Shortfall SRM} through the dual set as
\begin{equation*}
R_{\ES}^{\alpha }(X) = \sup_{\mu\in\mathcalboon{P}^{\tauare,\alpha }}\int_{\ba_{1,+}} \E{q}{-X}\,\dd\mu(q)
\end{equation*}
where \(\mathcalboon{P}^{\tauare,\alpha } = \left\lbrace\mu\in \Ptau\colon \mu \ll\mu_0 \text{ and } \dd\mu/\dd\mu_0\leq\frac{1}{\alpha }\right\rbrace\).

By \Cref{crl:dualSRM}, this SRM is coherent and \strict{} lower semicontinuous.
We have that $\mathcalboon{P}^{\tauare,\alpha }$ is weak$^*$-compact. Indeed, writing $\dd\mu = f\,\dd\mu_{0}$, this set is the
image of $\{f\in L^\infty(\mu_{0}):0\le f\le \frac{1}{\alpha },\ \int_{\ba_{1,+}} f\,\dd\mu_{0} = 1\}$,
which is weak$^*$-compact in $L^\infty(\mu_{0}) = L^1(\mu_{0})^*$. The map
$f\mapsto f\mu_{0}$ is weak$^*$-continuous, since $f\mapsto\int_U gf\,\dd\mu_{0}$ is
$\sigma(L^\infty,L^1)$-continuous for every $g\in C_b(U)$. Thus,
$\mathcalboon{P}^{\tauare,\alpha }$ is weak$^*$-compact, and the supremum is therefore attained. This SRM fails \wc{}. The same two-state structure gives a strict counterexample. Use again
$\Omega = \{1,2\}$, $q_1 = \delta_1$, $q_2 = \delta_2$, and
$\mu_0 = \frac12\delta_{q_1}+\frac12\delta_{q_2}$. Take $\alpha = 1$, so the density constraint forces $\mu = \mu_0$. Let
$x_1$ and $x_2$ be as above: $x_1(1) = -a$, $x_1(2) = 0$,
$x_2(1) = 0$, and $x_2(2) = -a$. For $X = \{x_1,x_2\}$, we have
$\widehat{-X}(q_1) = \widehat{-X}(q_2) = a$, so $R_{\ES}^{1}(X) = a$. For either singleton,
$R_{\ES}^{1}(\singleton{x_1}) = R_{\ES}^{1}(\singleton{x_2}) = \frac{a}{2}$.
Thus $R_{\ES}^{1}(X)>\sup_{x\in X}R_{\ES}^{1}(\singleton{x})$.
This proves failure of \wc{}.
As with scalar risk measures on $L^\infty$, the family $R_{\ES}$ can be used as building blocks for a larger class of SRMs:
\begin{equation*}
R(X) = \sup\limits_{m\in\mathcal{M}}\int_{(0,1]}R_{\ES}^{\alpha }(X)\,\dd m(\alpha),
\end{equation*}
where $\mathcal{M}$ is a collection of Borel probability measures on $(0,1]$.
Studying law invariance for random sets, although essential for Kusuoka-type representations, lies beyond the scope of this work.
\end{Exm}

\begin{Exm}[Aumann-type integral]
A natural example of an SRM is inspired by the \textit{Aumann integral}; see \citet{aumann1965integrals} for details. We define the \defin{Aumann-type integral of $X\in\CB$ with respect to $p$} as the subset of $\R$ given by
\begin{equation*}
\int_\Omega X\,\dd p = \left\lbrace\int_\Omega x\,\dd p\colon x\in X\right\rbrace.
\end{equation*}
From the Aumann-type integral, we can define an SRM through
\begin{equation*}
R(X) = -\inf\bigcup_{q\in Q}\int_\Omega X\,\dd q,\quad X\in\CB
\end{equation*}
where $Q\subseteq P^{\text{ac}}$ is non-empty, closed, and convex. $R$ is a well-defined SRM, since $L^\infty(p)\subseteq L^\infty(q)$ for any $q\in P^{\text{ac}}$, and it is coherent by the standard argument using monotonicity, additivity for constants, and positive homogeneity of both the integral and the infimum, together with the facts that the integral is linear and the infimum is superadditive. A case of particular interest is obtained by taking $Q = \singleton{p}$. Of course, many other real-valued operators over $\int_\Omega X\,\dd p$ can be considered beyond the infimum.
\end{Exm}

\begin{Exm}[New SRMs from old SRMs]
Given a collection $R_1,\dots,R_n$ of \emph{base} SRMs, it is possible to obtain new SRMs by composition with a multivariate transformation $g\colon \R^n\to\R$. The following constructions are useful. First, consider the \defin{risk-averse} $R_{\max}$ defined through
\begin{equation*}
R_{\max}(X) = \max\lbrace R_1(X),\dots,R_n(X)\rbrace,\qquad X\in\CB.
\end{equation*}
This map inherits the corresponding properties from $R_1,\dots,R_n$, namely
monotonicity, translation invariance, \Wc{}, convexity, positive
homogeneity, and the relevant lower semicontinuity properties whenever these hold
for each $R_i$. Moreover,
$\rvsetset{A}_{R_{\max}} = \bigcap_{i = 1}^n\rvsetset{A}_{R_i}$. In view of
\Cref{prp:SRM}, if the base SRMs are convex, then $R_{\max}^*$ is the lower
semicontinuous convex envelope of the pointwise minimum
$\min_{i\in\{1,\dots,n\}}R_i^*$. If the base SRMs are coherent, then
${\ba}_{R_{\max}}$ is the weak$^*$ closed convex hull of
$\bigcup_{i = 1}^n{\ba}_{R_i}$.
Alternatively, consider the \defin{permissive} SRM, $R_{\min}$, given by
\begin{equation*}
R_{\min}(X) = \min\lbrace R_1(X),\dots,R_n(X)\rbrace,\qquad X\in\CB.
\end{equation*}
In this case, we get $\rvsetset{A}_{R_{\min}} = \bigcup_{i = 1}^n\rvsetset{A}_{R_i}$, but note that $R_{\min}$ does not necessarily inherit convexity from $R_1,\dots,R_n$. Consequently, $R_{\min}$ need not admit a representation of the convex-duality form above. It also fails to inherit \wc{}, since we can easily take some $X\in\CB$ such that
\begin{equation*}
\min_{i = 1,\dots,n}\sup_{x\in X}R_i(\singleton{x})>\sup_{x\in X}\min_{i = 1,\dots,n}R_i(\singleton{x}) = \sup_{x\in X}R_{\min}(\singleton{x}).
\end{equation*}
Finally, one may consider an averaging of the form
\begin{equation*}
\bar{R}(X) = \sum_{i = 1}^n \lambda_i R_i(X),\qquad X\in\CB,
\end{equation*}
where $\lambda_1,\dots,\lambda_n\in[0,1]$ are convex weights such that $\sum_{i = 1}^n\lambda_i = 1$. This map inherits from the base SRMs all the properties of \Cref{def:risk}, with the exception of \wc{}, since we can take some $X\in\CB$ such that
\begin{equation*}
\sum_{i = 1}^n\lambda_i \sup_{x\in X}R_i(\singleton{x})>\sup_{x\in X}\sum_{i = 1}^n\lambda_i R_i(\singleton{x}) = \sup_{x\in X}\bar{R}(\singleton{x}).
\end{equation*}
The three constructions above can be viewed as ways to model uncertainty about the adequacy or quality of the risk assessments offered by each individual base SRM $R_1,\dots,R_n$.
\end{Exm}

\begin{Exm}[Knightian uncertainty]\label{ex: uncertainty sets}
This example offers an alternative way to tackle uncertainty.
Model misspecification and Knightian uncertainty in dynamic decision-making are widely studied frameworks for dealing with imperfect information and with the consequences of decisions made under ambiguity. To give an example, \citet{moresco2023uncertainty} consider a worst-case SRM similar to \eqref{eq:sup}.
In the context of financial losses represented by $x$, uncertainty may arise as the agent faces ambiguity about whether $x$ accurately models the true payoff. In such instances, it becomes prudent for the agent to consider a set of alternative random variables. These alternatives are payoffs that are considered ``close'' to $x$ or share common attributes, such as similar distributional features. The uncertainty set of $x$ is then written as $u(x)$, where $u\colon L^\infty \to \CB$ is translation invariant, positive homogeneous, convex, and monotone, i.e.\ if $x \leq y$, then $u(x) = : X \leq Y \coloneqq u(y)$. A large literature considers uncertainty sets, including uncertainty sets defined via, e.g., mixtures of distributions \citep{zhu2009worst}, moment constraints \citep{natarajan2009constructing}, divergence constraints \citep{wang2016likelihood}, and combinations of moment and divergence constraints \citep{bernard2022robust}. These are prime examples of uncertainty sets that are constructed as neighborhoods around a reference distribution or random variable, with radius given by a tolerance distance, that is
\begin{equation*}
u_D(x) \coloneqq \lbrace y\in L^\infty\colon D(x,y) \leq \varepsilon\rbrace,
\end{equation*}
where $D$ can be, for example, a seminorm \citep{gotoh2013robust}, the Wasserstein distance \citep{pflug2007ambiguity}, Kullback-Leibler (KL) divergence \citep{calafiore2007ambiguous}, general $f$-divergences, and expected scores \citep{moresco2023uncertainty}. Given a traditional coherent risk measure $\rho\colon L^\infty \to \R$ and an uncertainty set $u$, the corresponding \defin{worst-case robust} risk measure $\rho_u$ is given by
\begin{equation*}
\rho_u(x) = \sup_{y\in u(x)}\rho(y) = R_\rho(u(x)) = \E{Q^\rho}{-u(x)},\qquad x\in L^\infty,
\end{equation*}
where $R_\rho$ is the worst-case SRM associated with $\rho$, and
\begin{equation*}
Q^\rho = \lbrace q\in \ba_{1,+}\colon \sup_{x\in A_\rho} \E{q}{-x} <\infty\rbrace.
\end{equation*}

\end{Exm}

\begin{Exm}[Systemic risk]\label{example systemic risk}
The next example connects SRMs with systemic risk measures, as studied in \citet{Chen2013sys} and \citet{kromer2016systemic}, for instance.
Let an aggregation function be a surjective map $\lambda \colon\CB\to L^\infty$ that is monotone, translation invariant, normalized, concave, and continuous. A prominent example is $\lambda (X) = \essinf X$. Given a risk measure $\rho$, define an SRM by
\begin{equation*}
R(X)\coloneqq R_{\rho,\lambda}(X) = \inf\left\lbrace \alpha \in\R \colon \lambda (X)+\alpha \in A_\rho\right\rbrace,\qquad X\in\CB.
\end{equation*}
It is straightforward to verify that this map is a convex SRM. It is monetary or coherent under the corresponding assumptions on both $\lambda$ and $\rho$. Equivalently, by translation invariance,
\begin{equation*}
R(X)\coloneqq R_{A,\lambda}(X) = \inf\left\lbrace \alpha \in\R \colon \lambda (X+\singleton{\alpha })\in A\right\rbrace,\qquad X\in\CB,
\end{equation*}
where $A$ is a convex acceptance set. Both maps coincide when $A = A_\rho$ (equivalently, $\rho = \rho_A$).
For $\lambda (X) = \essinf X$, we obtain $R(X) = -\inf_{x\in X}\essinf x = \sup_{x\in X}\ML(x)$, where $\ML$ stands for the \defin{maximum loss} risk measure.

\end{Exm}

\begin{Exm}[Preferences]
Risk measures are closely linked to preferences. A traditional risk measure $\rho\colon L^\infty \to \R$ induces a preference relation $x \;\bar{\leq}\;y$ if and only if $\rho(x) \geq \rho(y)$. For details, see, for example, \citet{drapeau2013risk}. Following this idea, our SRM also induces a preference relation over $\CB$, which can be read as one basket of goods being preferred to another. More specifically, we define the preference order $\;\dot{\leq}\;$ by setting $X\;\dot{\leq}\;Y$ if and only if $R(X) \geq R(Y)$. Several studies address representations of preferences over sets of goods; see, for instance, \citet{kreps1979representation,dekel2001representing,dekel2007representing}. These works often rely on the concept of a ``subjective state space'', which can be challenging to identify explicitly. By Lemma 1 and Theorem 3 of \citet{dekel2001representing}, if $R$ is a continuous convex SRM, then it satisfies set-convexity and admits a representation of the form
\begin{equation}
\label{eq:utility-representation} R(-X) = u \left(\big(\sup\nolimits_{x \in X}\mathfrak{u}(q,x)\big)_{q \in S}\right),
\end{equation}
where $S$ is a state space, $\mathfrak{u}\colon S \times L^\infty \to \R$ is a state-dependent expected utility function that is affine in the second argument, and $u$ is an increasing aggregator function. In light of \Cref{thm:dualSRM}, we can take $S = \ba_{1,+}$,
\begin{equation*}
\mathfrak{u}(q,x) = \E{q}{x}, \quad \text{and} \quad u(\psi) = \sup_{\mu\in\Ptau}\left\lbrace\int_{\ba_{1,+}} \psi(q)\,\dd\mu(q)-R^*(\mu)\right\rbrace,
\end{equation*}
for admissible $\psi$; \cref{eq:utility-representation} is recovered by taking $\psi(q) = \sup_{x\in X}\mathfrak{u}(q,x)$, that is, $\psi = \widehat{X}\big|_S$. Under the attainment condition in \Cref{thm:dualSRM}, the supremum may be replaced by a maximum. The idea here is that the agent chooses a basket of assets (say, $X$), and then evaluates the utility of this choice when the ``latent state'' $q$ is revealed. A preference order $\bar{\leq}$ on the space of random variables admits a representation via an affine risk measure $\rho$ if and only if it satisfies the independence and Archimedean properties \citep[Proposition 1]{drapeau2013risk}. This result extends to the preference order $\dot{\leq}$ on sets of random variables, as established in Theorem 4 of \citet{dekel2001representing} and Theorem S2 of \citet{dekel2007representing}. Specifically, a preference order $\dot{\leq}$ over sets of random variables admits a representation via an affine SRM $R$ if and only if it satisfies the independence and Archimedean properties. In this case, there exists a measure $\mu \in \Ptau$ such that $ R(X) = \int_{\ba_{1,+}} \E{q}{-X}\,\dd\mu(q) - R^*(\mu). $
\end{Exm}

\section{Concluding remarks}

In this paper, we have established the theoretical foundations for set risk measures (SRMs), a functional-analytic framework that extends the domain of risk assessment from individual random variables to non-empty, closed, and bounded sets of financial positions. Unlike vector-valued risk measures---which typically map vectors to sets of eligible portfolios---SRMs operate as ``set-to-scalar'' maps. This approach captures the financial intuition of determining the minimal capital required to render an entire set of uncertain positions acceptable, effectively quantifying the risk of flexibility and ambiguity.

Our investigation required overcoming significant topological challenges inherent to hyperspaces. By equipping $\CB$ with the Hausdorff metric, we constructed an axiomatic scheme in \Cref{sec:SRM} that adapts classical properties---such as monotonicity, translation invariance, and convexity---to the arithmetic of sets. A primary technical contribution of this work, detailed in \Cref{sec:dual}, is the dual representation theorem for convex SRMs. By embedding the domain into the space of bounded continuous functions on the dual unit ball, $C_b(U)$, we derived a representation involving regular, $\tau$-additive (for nets) probability measures. Crucially, we identified the role of the \strict{} topology in this embedding, providing a robust analytical tool for future developments in set-valued analysis.

In \Cref{sec:WC}, we further explored the subclass of worst-case SRMs, which aligns with the paradigm of robust risk management. We demonstrated that under the property of WC-boundedness, an SRM can be decomposed into the supremum of risks of individual elements. This connects our framework to existing literature on ambiguity aversion and Knightian uncertainty, providing a rigorous bridge between set-based risk and model uncertainty.

The versatility of the SRM framework, illustrated by the examples in \Cref{sec:examp}, opens several avenues for future research. First, while we focused on static measures, extending SRMs to a dynamic setting is a natural next step, especially in connection with time-consistency and recursive acceptance sets. Second, law invariance and comonotonic additivity for random sets remain to be fully characterized. Establishing Kusuoka-type representations in this context would be valuable for standardizing SRMs in regulatory applications. Finally, exploring algebraic operations beyond the convex hull---such as set differences or complements---could yield insights into risk budgeting and risk transfer problems where the removal of specific assets from a portfolio is the primary concern.

\bibliography{references}
\bibliographystyle{agsm}
\end{document}